\documentclass[journal,  onecolumn, 12pt,  draftclsnofoot,]{IEEEtran}
\usepackage{graphicx,float}
\usepackage{subfigure}
\usepackage{array}
\usepackage{cite}

\usepackage{bm}
\usepackage{amsmath}
\usepackage{amsfonts}
\usepackage{amssymb}
\usepackage{algorithmic}
\usepackage[ruled]{algorithm2e}
 
\usepackage{xcolor}

\graphicspath{{figure/}}
\usepackage{amsthm}
\newtheorem{lemma}{Lemma}
\newtheoremstyle{noparens}%
  {}% space above
  {}% space below
  {}% body font \itshape
  {}% indent amount
  {\itshape}% theorem head font \bfseries
  {:}% punctuation after theorem head
  { }% space after theorem head
  {\thmname{#1}\thmnumber{ #2}\mdseries\thmnote{ #3}}% theorem head spec (can be left empty, meaning 'normal')
\theoremstyle{noparens}
\newtheorem{lemmaNoParens}[lemma]{Lemma}

\newcommand{\BE}{\begin{equation}}
\newcommand{\EE}{\end{equation}}
\newcommand{\BS}{\begin{subequations}}
\newcommand{\ES}{\end{subequations}}
\renewcommand{\bf}{\bm}


\begin{document}

\title{RIS-Aided Multiuser MIMO-OFDM with Linear Precoding and Iterative Detection: Analysis and Optimization}

\author{Mingyang Yue,
        Lei Liu,~\IEEEmembership{Member,~IEEE},
        and Xiaojun Yuan,~\IEEEmembership{Senior Member,~IEEE}
        
        \thanks{Mingyang Yue and Xiaojun Yuan are with the National  Key Laboratory of Science and Technology on Communications, University of Electronic Science and Technology of China, Chengdu 611731, China (e-mail:myyue@std.uestc.edu.cn; xjyuan@uestc.edu.cn).}
        \thanks{Lei Liu is with the School of Information Science, Japan Advanced Institute of Science and Technology, Nomi 923-1292, Japan (e-mail: leiliu@jaist.ac.jp).}
        }

\maketitle

\begin{abstract}
In this paper, we consider a reconfigurable intelligence surface (RIS) aided uplink multiuser multi-input multi-output (MIMO) orthogonal frequency division multiplexing (OFDM) system, where the receiver is assumed to conduct low-complexity  iterative detection. We aim to minimize the total transmit power by jointly designing the precoder of the transmitter and the passive beamforming of the RIS. This problem can be tackled from the perspective of information theory. But this information-theoretic approach may involve prohibitively high complexity since the number of rate constraints that specify the capacity region of the uplink multiuser channel is exponential in the number of users. 
To avoid this difficulty, we formulate the design problem of the iterative receiver under the constraints of a maximal iteration number and  target bit error rates of users. To tackle  this challenging problem, we propose a groupwise successive interference cancellation (SIC) optimization approach, where the signals of users are decoded and cancelled in a group-by-group manner.  We present a  heuristic user grouping strategy, and  resort to the alternating optimization technique to iteratively solve the precoding and passive beamforming sub-problems. Specifically, for the  precoding sub-problem, we employ  fractional programming to convert it to a convex problem; for the passive beamforming sub-problem,  we  adopt successive convex approximation  to deal with the unit-modulus constraints of the RIS.   We show that  the proposed groupwise SIC  approach has significant advantages in both performance and computational complexity, as compared with the counterpart approaches.
\end{abstract}
\newpage
\begin{IEEEkeywords}
Reconfigure intelligent surface, MIMO-OFDM, precoding, passive beamforming, iterative detection
\end{IEEEkeywords}

\IEEEpeerreviewmaketitle

\section{Introduction} \label{sec-intro}
Reconfigurable intelligent surface (RIS) has been regarded as a key enabling technology for the sixth-generation (6G) wireless communications\cite{6G}. RIS is composed of a large number of low-cost passive reflecting elements, where a controller is equipped to adjust the reflection coefficient of each element in real time. By effectively designing the passive beamforming of the elements, RIS can  reconfigure the wireless propagation environment by intelligently manipulating the reflection direction of the incident electromagnetic wave. In addition,  RIS also has the advantages of flexible deployment,  low energy consumption, and low noise \cite{zhangrui1}. 
Extensive research has been conducted on the design of RIS   to improve the system performance in terms of energy and spectral efficiency \cite{tradeoff,se1,ee1,se2,se3} and channel capacity \cite{capacity, capacity2, capacity3}.

 The optimization of  RIS  to improve the performance of multiuser systems has been investigated, e.g., in \cite{tradeoff,  ee1, yanggang, SOCP, shiyuanming,sum-rate2,sum-rate4,sum-rate1, sum-rate5,sum-rate3}. Particularly,  the authors in \cite{yanggang} considered a RIS-aided downlink multiuser system, and  designed the active and passive beamforming to maximize the minimal user rate. Ref. \cite{SOCP} aimed to minimize the total transmit power under the user rate constraints  for the RIS-aided downlink multiuser multiple-input-single-output (MISO) system. Ref. \cite{shiyuanming} proposed a difference-of-convex  algorithm to jointly design   active and passive beamforming by minimizing the total transmit power of all  users. The (weighted) sum rate of the multiuser system has been studied  in \cite{sum-rate2} and \cite{ sum-rate4}, and has been extended to other scenarios, such as millimeter wave \cite{sum-rate1, sum-rate5} and multi-RIS-aided cooperative transmission \cite{sum-rate3}. 
Furthermore, the energy efficiency, defined as the sum rate normalized by the energy consumption, is investigated for RIS aided  downlink  systems \cite{se1} and multiple-input-multiple-output (MIMO)  uplink systems \cite{tradeoff,ee1}. 

% In this paper, different from above works, we assume that the rate of each user  is fixed, and formulate the optimization problem to minimize the total transmit power by jointly optimizing the precoding of transmitter and  the passive beamforming under the constraints of capacity region, for the RIS aided MIMO NOMA uplink systems. Due to the unit-modulus constraint of  RIS elements, the optimization is non-convex.  We propose an optimization approach, called information-theoretic approach, which resorts to the alternating optimization and relaxes the unit-modulus constraint to obtain an approximate solution. However, this approach has a huge computational complexity when optimizing, which increases exponentially with the number of users. 

The above works are  designed from the perspective of information-theoretic performance metrics. In other words, these works assumed an ideal receiver that can achieve the capacity of the system. However, it is so-far unclear how to design a practical low-complexity  receiver to achieve the capacity of the RIS-aided multiuser MIMO channel. In fact, a practical low-complexity receiver may perform quite far away from the capacity, and the capacity-based optimizations may not achieve the expected performance in a practical system with limited computational capability. Therefore,  it is desirable to optimize the performance of the RIS-aided system based on more practical performance metrics rather than on information-theoretic metrics.

In this paper, we study the design of the RIS-aided uplink multiuser MIMO orthogonal frequency division multiplexing (OFDM) system equipped with a practical low-complexity transceiver. Specifically, we adopt the  low-complexity receiver with iterative linear minimum-mean-square-error (LMMSE) detection and decoding,  which has been previously studied in traditional communication systems \cite{evolution, precoder, achieve-rate1,achieve-rate2}. In \cite{evolution}, the authors proposed the low-complexity iterative detection framework, and established a   signal-to-interference-plus-noise ratio (SINR) variance state evolution technique to characterize the performance. Ref. \cite{precoder} studied the linear precoder design for the iterative receiver. In \cite{achieve-rate1}, the authors discussed the achievable rate in a single-user system based on the iterative receiver, and showed that the water-filling channel capacity  can be achieved by  appropriate precoder design. Furthermore,  \cite{achieve-rate2} showed that for the MIMO-NOMA system  this  receiver can achieve the sum capacity with appropriately designed channel codes. 
 
 We aim to jointly design the precoder of transmitter and the passive beamforming of RIS, to minimize the total transmit power for the uplink multiuser MIMO-OFDM system.  This optimization problem can be solved by  extending the  approach in \cite{capa} to the multiuser scenario. Yet, the number of rate constraints that specifies the capacity region of the  multiuser access channel (MAC) channel is exponential in the number of users. Thus, this information-theoretic  approach will cause a prohibitively high complexity in practice, even for a MAC channel with a moderate number of users.
 
 To address this issue,  we  formulate the joint optimization problem for the iterative LMMSE receiver under the constraints of maximal iteration number and allowed bit error rate (BER).  The path condition  for BER constraint  is derived via the state evolution (SE) \cite{evolution}, where a feasible path needs to be found in a high-dimension variance space.
 This  optimization  is  challenging even for a  traditional multiuser MIMO-OFDM systems without RIS \cite{Convergence}, let alone  the handling of the extra non-convex constraints of the RIS phase adjustments. 
 To tackle this  problem,  we propose a group successive interference cancellation (SIC)  approach, where  users are decoded  and cancelled in a group-by-group manner.  We first fix the user grouping, and  resort to the alternating optimization for  precoding and passive beamforming.  Specifically, we convert the precoding sub-problem to a convex problem by using the fractional programming method \cite{FP}. For the sub-problem of passive beamforming,   we  adopt the successive convex approximation  to deal with the unit-modulus constraints of RIS. Besides,  we  propose a heuristic low-complexity user grouping strategy to avoid the exhaustive search over all possible user groupings. Numerical results show that the proposed groupwise SIC  approach substantially  outperforms the counterparts, especially when the allowed iteration number of the receiver is relatively small.
 The contributions of this work are summarised as follows:
\begin{itemize}
    \item   We  investigate the joint precoding and passive beamforming design to  minimize the total transmit power of the RIS-aided uplink multiuser MIMO-OFDM system. In contrast to the  existing works, we consider the system optimization for the low-complexity iterative LMMSE receiver.   We establish the state evolution to characterize the system performance, and formulate the joint optimization problem for the iterative receiver under the constraints of target BERs and a maximal iteration number . 
  
    \item We propose a  groupwise SIC  approach to solve the challenging optimization problem. We employ a heuristic but efficient user grouping strategy to avoid the exhaustive search over all possible user groupings.  For a fixed user grouping, we  tackle the problem by alternately solving the  precoding and passive beamforming sub-problems. We employ the fractional programming method to convert the precoding sub-problem to a convex problem. We further show that the passive beamforming sub-problem  is a feasibility-check problem, which can be solved by skilfully expanding the feasible region and applying  successive convex approximation. 
   
    \item   We show that the complexity of the groupwise SIC  approach is much lower than that of information-theoretic  approach. We also show the groupwise SIC  approach substantially  outperforms  the information-theoretic  approach and other counterparts, especially for a relatively small allowed iteration number of the receiver.
    % We  find that when the number of receiver antennas is large enough, such as more than 8,  the power allocation between subcarriers  in precoding of groupwise SIC  approach  is getting identical to each other. 
\end{itemize}

The rest of this paper is organized as follows. Section \ref{sec-system} presents the channel model and the transceiver for a RIS-aided uplink MIMO-NOMA system. In Section \ref{sec-problem},  the state evolution is described, and the optimization problem for the iterative receiver is formulated. In Section \ref{sec-sic}  the groupwise SIC  approach is described.  Numerical results are provided in Section \ref{sec-simu}. Section \ref{sec-con} concludes the paper.

\emph{Notions:} We use a bold symbol lowercase letter and bold symbol capital letter to denote a vector and a matrix, respectively. 
The Frobenius norm, trace, transpose, conjugate transpose, and inverse of a matrix are denoted by $\|\cdot\|_{\text{F}}$, $\text{tr}(\cdot)$, $(\cdot)^{\text{T}}$, $(\cdot)^{\text{H}}$ and $(\cdot)^{-1}$, respectively; $[\boldsymbol{A}]^2$ denotes $\boldsymbol{A}\boldsymbol{A}^{\text{H}}$;
 $\text{diag}(\boldsymbol{a})$ forms a diagonal matrix with the diagonal elements in $\boldsymbol{a}$;
  $\|\cdot\|_2$ denotes $\ell_2$-norm of a vector; 
 $|\cdot|$ denotes the modulus of a complex number; 
 the complex conjugate of a vector is denoted by $(\cdot)^*$;
$\otimes$ denotes the Kronecker product;
$\odot$ denotes the Hadmard product;
$\ast$ denotes the circular convolution;
$\nabla$ denotes the gradient operator;
$\mathcal{O}$ is the big-O  notation.
The complex Gaussian distribution of $\boldsymbol{x}$ with mean
$\boldsymbol{m}$ and covariance matrix $\boldsymbol{\Sigma}$ is denoted by   $\boldsymbol{x}\sim {\cal{CN}}(\boldsymbol{m},\boldsymbol{\Sigma})$.

\begin{figure}[b!]
 \centering
    \includegraphics[width=3in]{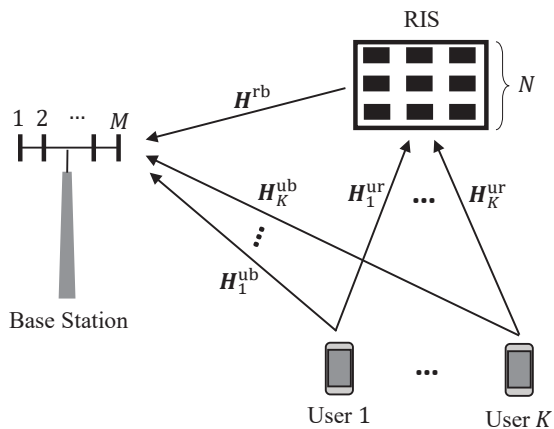}
    \caption{An uplink multiuser  MIMO-OFDM system consisting of $K$ single-antenna users  and an $M$-antenna  base station, where a RIS with $N$ elements is deployed to enhance the communication. The channels from user $k$ to the RIS, from user $k$ to the base station and from the RIS to the base station are denoted by $\bm{H}_{k}^{\text{ur}}$, $\bm{H}_{k}^{\text{ub}}$ and $\bm{H}^{\text{rb}}$, respectively.} 
    \label{Figsystem}
\end{figure}

\section{System Model} \label{sec-system}

\subsection{Channel Model}
  As shown in Fig. \ref{Figsystem}, we consider an uplink RIS-aided multiuser  MIMO-OFDM system, where a RIS with $N$ reflecting elements is deployed to enhance the communication between $K$ single-antenna users and an $M$-antenna base station (BS). Let $J$ denote the number of subcarriers of each OFDM symbol, and these $J$ subcarriers are shared by all the users. 
  We assume frequency-selective fading channels, and the user-BS, user-RIS and RIS-BS links contain $L_{\text{ub}}$, $L_{\text{ur}}$ and $L_{\text{rb}}$ taps in the impulse response, respectively\cite{ISI}. Each user adopts a cyclic prefix (CP) of length $L_{\text {cp}}$ with $L_{\text {cp}} \ge \text{max}\{L_{\text{ub}}, L_{\text{ur}}+L_{\text{rb}}\} $  to eliminate  inter-symbol interference.  
  Then,   the  received baseband  signal  in the time domain at the $m$th antenna after   CP removal is given by
    \begin{equation}\label{r-conv}
    \bm{r}_m = \sum_{k=1}^{K} \left(\bm{h}^{\text{ub}}_{k,m}  + \sum_{n=1}^{N} \theta_n \bm{h}_{k,m,n}^{\text{rb}} \ast   \bm{h}_{k,n}^{\text{ur}}   \right)\ast \bm{s}_k+ \bm{n}_m,
  \end{equation}
  where $\bm{s}_k \in \mathbb{C}^{J \times 1}$  is the transmit signal vector in the time domain, 
  $\theta_n$ the reflecting  coefficients of the $n$th RIS element, 
  $\bm{h}^{\text {ub}}_{k,m} \in \mathbb{C}^{J \times 1}$ the channel  from  user $k$ to the BS's $m$th antenna, 
  $\bm{h}_{k,n}^{\text{ur}}\in \mathbb{C}^{J \times 1}$  the channel from  user $k$ to the $n$th RIS element,
  $\bm{h}^{\text{rb}}_{m,n}\in \mathbb{C}^{J \times 1}$ the channel  from the  $n$th RIS element to the BS's $m$th antenna, and 
  $\bm{n}_m$  the additive white Gaussian noise (AWGN). 
   For simplicity, we assume that  each $\theta_n$ is constant within the considered frequency band. Besides, the phase can be adjusted independently in $[0,2\pi]$, and all the reflecting  coefficients are unit-modulus, i.e., $|\theta_n|=1, \forall n$.  We also assume  that  perfect channel state information (CSI) is available. In practice, the CSI can be acquired by existing channel estimation methods; see, e.g., \cite{ce} and \cite{ce2}.

With \eqref{r-conv}, the time-domain  channel matrix of the direct link between user $k$ and the BS is a block-circulant matrix given by 
  \begin{align} \label{H^{ub}}
  \bm{H}_k^{\text{ub}} & \triangleq
  \begin{bmatrix}
  \bm{H}_k^{\text{ub}}(1)       & \bm{H}_k^{\text{ub}}(J) & \cdots & \bm{H}_k^{\text{ub}}(2)  \\
  \bm{H}_k^{\text{ub}}(2)      &      \bm{H}_k^{\text{ub}}(1)  & \cdots    & \bm{H}_k^{\text{ub}}(3) \\
  \vdots         & \vdots    & \ddots    & \vdots      \\
  \bm{H}_k^{\text{ub}}(J)     & \bm{H}_k^{\text{ub}}(J-1) & \cdots  & \bm{H}_k^{\text{ub}}(1)  \\
  \end{bmatrix} \in {\mathbb{C}}^{JM\times J} ,
\end{align}
  where  each block  $\bm{H}_k^{\text{ub}}(j)=[ h_{k,1}^{\text{ub}}(j),h_{k,2}^{\text{ub}}(j),\ldots,h_{k,M}^{\text{ub}}(j) ]^{\text{T}} \in {\mathbb{C}}^{M\times 1}$ is the channel  at the $j$th tap. 
  Similarly, the  time-domain  channel matrix  between the $k$th user and  the $n$th RIS element  $\bm{H}^{\text{ur}}_{k,n} \in \mathbb{C}^{J \times J}$  is a circulant matrix,   and the  time-domain  channel matrix  between  the $n$th RIS element and the BS $\bm{H}^{\text{rb}}_{n} \in \mathbb{C}^{JM \times J}$ is a block-circulant matrix. 
  Let 
  \begin{align}
     &\bm{r}=\left[\bm{r}^{\text{T}}(1),\bm{r}^{\text{T}}(2),\ldots,\bm{r}^{\text{T}}(J)\right]^{\text{T}} \in {\cal{C}}^{JM\times 1}, \nonumber\; \qquad \bm{r}(j)=\left[r_1(j),r_2(j),\ldots,r_{M}(j)\right]^{\text{T}},\\
     &\bm{n}=\left[\bm{n}^{\text{T}}(1),\bm{n}^{\text{T}}(2),\ldots,\bm{n}^{\text{T}}(j)\right]^{\text{T}} \in {\cal{C}}^{JM\times 1},  \qquad\bm{n}(j)=[{n}_1(j),n_2(j),\ldots, n_{M}(j)]^{\text{T}}.\nonumber
  \end{align} 
  Then,  the received baseband signal  is given by
  \begin{equation}\label{r-mat}
    \bm{r} = \sum_{k=1}^{K} \left(\bm{H}^{\text{ub}}_k  + \sum_{n=1}^{N}\theta_n \bm{H}_{k,n}^{\text{rb}} \bm{H}_{k,n}^{\text{ur}} \right) \bm{s}_k + \bm{n}.
  \end{equation}
 The frequency-domain channel matrix $\bm{G}_k^{\text{ub}}$ corresponding to  $\bm{H}^{\text{ub}}_k$ is a block-diagonal matrix given by
 \begin{subequations}\label{G}
    \begin{align} \label{Gk}
  \bm{G}_k^{\text{ub}} & =
  \begin{bmatrix}
  \bm{G}_k^{\text{ub}}(1)  &  \bm{0}    & \cdots & \bm{0} \\
     \bm{0} & \bm{G}_k^{\text{ub}}(2)  & \ddots     & \vdots \\
    \vdots &  \ddots     &  \ddots    &   \bm{0}  \\
   \bm{0}&  \bm{0}& \cdots  & \bm{G}_k^{\text{ub}}(J)  \\
  \end{bmatrix} \in {\mathbb{C}}^{JM\times J} 
\end{align}
with its $j$th diagonal block calculated as
  \begin{equation}\label{g}
    \bm{G}^{\text{ub}}_k(j)=  \frac{1}{J}\sum_{l=1}^{J} \bm{H}_{k}^{\text{ub}}(l)\text{exp}(-i2\pi j(l-1)/J),
 \end{equation}
 \end{subequations}
  where $i= \sqrt{-1}$. A concise form of \eqref{G} is given by  
     \begin{align}\label{G1}
    \bm{G}^{\text{ub}}_k= \bm{F}_M  \bm{H}^{\text{ub}}_k \bm{F}^{\text{H}}, \qquad\qquad
    \bm{H}^{\text{ub}}_k= \bm{F}_M^{\text{H}}  \bm{G}^{\text{ub}}_k \bm{F}, 
    \end{align}  
  where $\bm{F} \in \mathbb{C}^{J\times J}$ is the $J$-by-$J$ unitary discrete Fourier transform (DFT) matrix, $\bm{F}_M =\bm{F} \otimes \bm{I}_M$ with $\bm{I}_M$ being the $M\times M$ identity matrix. Similar to $\bm{G}^{\text{ub}}_k$,  $\bm{G}_{n}^{\text{ur}} \in \mathbb{C}^{JM \times J}$ is a block-diagonal matrix with block-size $M$-by-1, and  $\bm{G}_{k,n}^{\text{rb}} \in \mathbb{C}^{J \times J}$ is a diagonal matrix. After applying the DFT at the BS, we obtain the frequency-domain received signal as
      \begin{align}\label{r-fre}
    \bm{r}' &= \bm{F}_{M}\bm{r} 
    =\sum_{k=1}^{K} \left(\bm{G}^{\text{ub}}_k  + \sum_{n=1}^{N}\theta_n \bm{G}_{k,n}^{\text{rb}} \bm{G}_{n}^{\text{ur}} \right) \bm{F}\bm{s}_k + \bm{\eta},
  \end{align}
  where $\bm{\eta}=\bm{F}_{M}\bm{n} \sim \mathcal{CN}(\bm{0}, \sigma^2\bm{I})$ is an AWGN.
 
  \begin{figure}[t]
    \centering
    \includegraphics[width=5in]{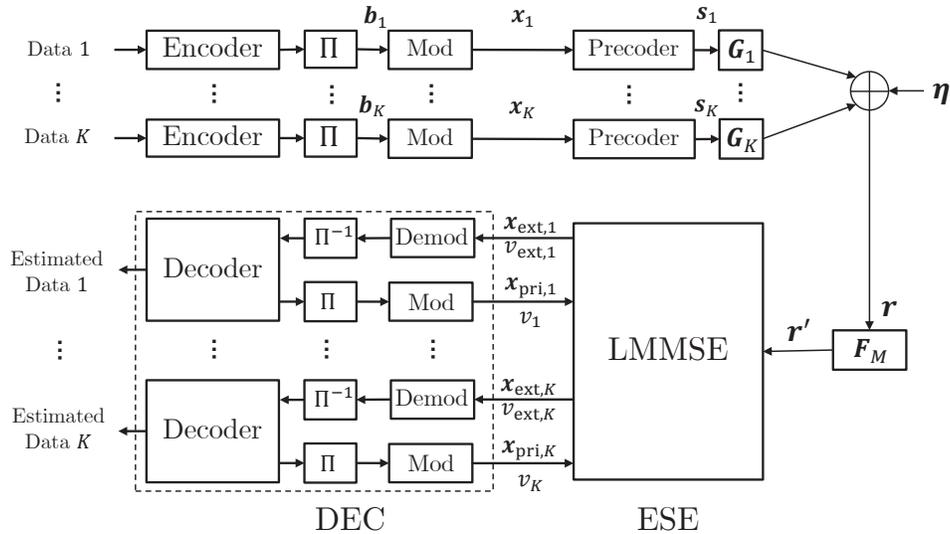}
    \caption{The transmitter and (iterative) receiver structure for the uplink MIMO-NOMA system. The transmitter consists of an FEC encoder, an interleaver (denoted by $\Pi$), a modulator (denoted by Mod) and a linear precoder.  The iterative receiver consists of a LMMSE-ESE module and a DEC module that handle the linear constraint and the coding constraint, respectively.}
    \label{receiver}
    \end{figure}

\subsection{Transmitter Structure}

  The transmitter is illustrated at the upper part of Fig. \ref{receiver}. For each user $k$, its data is first encoded by forward-error-correction codes, such as the low-density parity-check (LDPC) code, with  coding rate $R_{k}$, and then permuted by an interleaver to obtain $\bm{b}_{k}=\left[\bm{b}_{k,1}^T, \bm{b}_{k,2}^T,\dots,\bm{b}_{k,J}^T\right]^T$, where  $\bm{b}_{k,j} \in \{0,1\}^{Q_{k}}$ and $Q_{k}$ is  the number of bits per modulated symbol of user $k$. For simplicity, we assume that all the users adopt the same type of modulation,  implying $Q_1=\ldots=Q_K=Q$.
  The modulated vector $\bm{x}_k \in \mathbb{C}^{J \times 1}$ is generated by mapping each $\bm{b}_{k,j}$ to a discrete   constellation $\mathcal{X}_k = \{c_{k,1}, c_{k,2}, \dots , c_{k,2^{Q}}\}$ with zero mean and unit variance. 
  Then, $\bm{x}_k$ is  linearly precoded  by \cite{precoder}
  \begin{equation}\label{precoding}
    \bm{s}_k= \bm{F}^{\text{H}} \bm{W}_k \bm{F} \bm{x}_k,
  \end{equation}
  where  $\bm{W}_k=\text{diag}\{W_k(1,1),\ldots,W_k(J, J)\}$ is used for power allocation among the subcarriers.  Substituting \eqref{precoding} into \eqref{r-fre}, we obtain
  \BS %   
  \begin{align} 
    \bm{r}' &=\sum_{k=1}^{K}\left(\bm{G}^{\text{ub}}_k  + \sum_{n=1}^{N}\theta_n \bm{G}_{n}^{\text{rb}} \bm{G}_{k,n}^{\text{ur}} \right) \bm{W}_k \bm{F} \bm{x}_k + \bm{\eta}\\
            &=  \sum_{k=1}^{K}\bm{G}_k(\bm{\theta}) \bm{W}_k \bm{F} \bm{x}_k + \bm{\eta}, \label{linear-const}
  \end{align}
  \ES
  where  $\bm{G}_k(\bm{\theta}) = \bm{G}^{\text{ub}}_k  +  \sum_{n=1}^{N}\theta_n \bm{G}_{n}^{\text{rb}} \bm{G}_{k,n}^{\text{ur}}$ is the equivalent frequency-domain channel matrix, and $\bm{\theta}=[\theta_1,\ldots,\theta_N]^{\text{T}}$.% $\bm{G}_{\text{r}} \in \mathbb{C} ^{J \times N}$ with ${G}^{\text{R}}_k(j,n) ={G}_n^{\text{ur}}(j,j){G}_n^{\text{rb}}(j,j)$, 

\subsection{Receiver Structure}
  The existing works on RIS mostly aim to improve the system performance by assuming an ideal receiver with capacity-achieving performance. Such a capacity-achieving receiver usually involves  prohibitively high complexity. Instead, we here consider a low-complexity iterative receiver \cite{precoder}, which is  meaningful for practical implementation.  
  
  As illustrated in the lower part of Fig. \ref{receiver},  the iterative receiver consists of an elementary signal estimator (ESE) that  handles the linear constraint in \eqref{linear-const},
  and a generalized decoding processor (DEC) that handles the coding constraint \cite{precoder}. The ESE and the DEC are executed iteratively. 
  \subsubsection{LMMSE-ESE Module}
 The ESE carries out the linear minimum mean-square error (LMMSE) estimation based on the channel input $\bf{r}'$ and the messages from the DEC $\{\bm{x}_{\text{pri},k},  v_{k}\}$. The  signal $\bm{x}_k$ is approximately modeled by a  Gaussian distribution  $\mathcal{CN}(\bm{x}_{\text{pri},k},  v_{k}\bm{I})$. Given $\bm{r}'$,  the \textit{a posterior} covariance matrix and \textit{a posterior} mean of $\bm{x}_k$ are expressed as \cite{precoder} 
\BS \label{lmmse}
\begin{align}
  \bm{V}_{\text{post},k} &= v_{k}\bm{I} - v_{k}^2 \bm{A}_{k}^{\text{H}} \bm{V}^{-1}\bm{A}_k,\\
  \bm{x}_{\text{post}, k}  &= \bm{x}_{\text{pri,k}} + v_{k} \bm{A}_{k}^{\text{H}}\bm{V}^{-1} \left( \bm{r}' -  \sum_{k'=1}^{K}\bm{A}_{k'}\bm{x}_{\text{pri},k'}\right),
\end{align}
 where 
 \begin{align}
  \bm{A}_{k} &= \bm{G}_k(\bm{\theta}) \bm{W}_k \bm{F}, \\
  \bm{V}  &= \sum_{k'=1}^{K}v_{\text{pri},k'}\bm{A}_{k'}\bm{A}_{k'}^{\text{H}}+\sigma^2\bm{I}.
\end{align}
\ES

Following \cite{single-sys}, 
% to reduce the correlation between the input and output of the LMMSE-ESE, 
we calculate the extrinsic variance and mean as the output of the LMMSE-ESE:
\BS \label{ext}
\begin{align}
  v_{\text{ext},k} &= \left( v_{\text{post},k}^{-1} -  v_{k}^{-1} \right)^{-1}, \label{vext}\\
  \bm{x}_{\text{ext},k} &= v_{\text{ext},k} \left(v_{\text{post},k}^{-1}{\bm{x}_{\text{post},k}} - v_{k}^{-1}{\bm{x}_{\text{pri},k}}\right),\label{xext}
\end{align}
\ES
where $v_{\text{post},k} = \frac{1}{J} \text{tr}\{\bm{V}_{\text{post},k}\}$. In  \eqref{ext}, $\bm{x}_{\text{ext},k}$ is modeled as  an AWGN observation of $\bm{x}_k$:
\begin{equation} 
  \bm{x}_{\text{ext},k} = \bm{x}_k + \bm{\xi}_k, \label{ext-channel}
\end{equation}
where $\bm{\xi}_k \sim \mathcal{CN}(\bm{0}, \rho^{-1}_k\bm{I})$ with $\rho_k=v_{\text{ext},k}^{-1}$ being the SINR of the effective channel in \eqref{ext-channel}.

\subsubsection{DEC Module} 
The  DEC module is composed of \textit{a posterior} probability (APP) decoders, de-interleavers/interleavers and soft demodulators/modulators for all the users.
Given $\bm{x}_{\text{ext},k}$, the soft demodulator calculates  the log-likelihood ratio (LLR)  of each $\bm{b}_{k,j}(q)$ as 
\begin{equation}
  \lambda_{k,j} (q) = \ln \frac{p(b_{k,j}(q)=0|x_{{\text{ext}}, k}(j))}{p(b_{k,j}(q)=1|x_{{\text{ext},k}}(j))},\; j=1,2,\ldots,J,q=1,2,\ldots,Q.
\end{equation}
% Then $\bm{\lambda}_k=\left[\bm{\lambda}_{k,1}^{\text{T}},\bm{\lambda}_{k,2}^{\text{T}}, \dots,\bm{\lambda}_{k,J}^{\text{T}}\right]^{\text{T}}$ is permuted by 
After  de-interleaving, APP decoding and interleaving,  the \textit{a posterior}  LLR for each $\bm{b}_{k,j}$ is obtained, denoted by $\bm{\gamma}_{k,j}$. Then we update the  mean and variance of each $x_{k,j}$:
\BS\begin{align}
    \bar{x}_{k,j} &= \text{E}\left[x_{k,j} |\bm{\gamma}_{k,j}\right] = \sum_{c_{k} \in \mathcal{X}_k} c_{k}  p \left( x_{k,j} =c_{k}|\bm{\gamma}_{k,j} \right), \\
     \bar{v}_k &= \frac{1}{J} \sum_{j=1}^{J} \sum_{c_{k} \in \mathcal{X}_k} |c_{k}-\bar{x}_{k,j}|^2  p \left( x_{k,j} =c_{k}|\bm{\gamma}_{k,j}\right).
\end{align}\ES
Similarly to \eqref{vext} and \eqref{xext}, we update the \textit{a priori} mean and variance of $\bm{x}_k$ for the LMMSE-ESE by 
\BS\begin{align} \label{vext2}
  v_{k} &= \left( {\bar{v}_k}^{-1} -  v_{\text{ext},k}^{-1} \right)^{-1}, \\%
  \bm{x}_{\text{pri},k} &= v_{k} \left({\bar{v}_k^{-1}}{\bm{\bar{x}}_k} - v_{\text{ext},k}^{-1}{\bm{x}_{\text{ext},k}}\right).%
\end{align}\ES 

The remainder of this paper is devoted to the optimization of the system performance over the power allocation matrix $\{\bm{W}\}_k$ and the RIS phase shifts $\bm{\theta}$. This optimization   can be done based on the well-known capacity region of the multiple access (MAC) channel, with the details given in Appendix \ref{sec-info}. However, there are two issues with this information-theoretic approach. First, the complexity of the formation-theoretic  approach is exponential in the number of users (i.e., $K$), since the capacity region of a $K$-user MAC channel involves $2^{K}-1$ rate constraints. Second, a practical iterative receiver may perform far away from the capacity due to implementation limitations. For example, with a limited computational power, the iterative receiver may strictly limit its iteration number in exchanging the messages between the ESE and the DEC. In this case, the optimization result based on information theory may be not a good choice for the practical system. For these reasons, we propose a state-evolution based optimization approach with lower complexity and better performance in the following section.

\section{Performance Analysis} \label{sec-problem}
In this section,  we describe the state evolution to characterize the performance of the iterative receiver, and then formulate the joint precoding and passive beamforming optimization problem. First, we provide the transfer functions of the DEC and the LMMSE-ESE, based on which the state evolution is established. Then,  we formulate the joint optimization problem to reduce the total transmit power   under the constraints of user BER and maximal iteration number. 

\subsection{State Evolution (SE)}  
 
\begin{figure}
    \centering
    \includegraphics[width=2in]{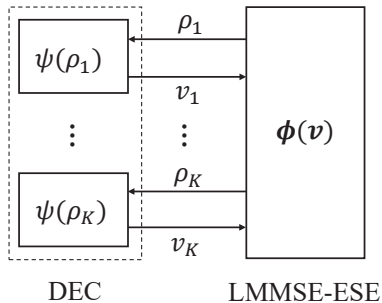}
    \caption{An illustration of the overall evolution process of the iterative receiver. The  LMMSE-ESE transfer function $\bm{\rho}=\bm{\phi}(\bm{v})$ is a vector function, and the DEC transfer function consists of $K$ separable functions  $ v_k= \psi(\rho_k), \forall k$.}
    \label{transfer}
\end{figure}
The SE is a semi-analytical method for performance evaluation of an iterative receiver. Specifically, we use the  output SINR $\bm{\rho}=\left[\rho_1,\ldots,\rho_K\right]^{\text{T}}$ to characterize the performance of the LMMSE-ESE, and the output variance $\bm{v}=\left[v_1,\ldots,v_K\right]^{\text{T}}$ to characterize the performance of the DEC, while the output of a module is the input of the other module.
As shown in Fig. \ref{transfer}, we track the  performance  by the following recursion: Start with $\bm{v}^{(0)}=[1,\ldots,1]^{\text{T}}$,
\BS\label{Eqn:SE}\begin{align}
  \bm{\rho}^{(t)}& =\bm{\phi}(\bm{v}^{(t-1)}),  \\
  \bm{v}^{(t)}& =\bm{\psi}(\bm{\rho}^{(t)}),
\end{align}\ES
where the superscript $t=1,2,\ldots,T$ is the iteration number with $T$ being the maximum iteration number, the MIMO function  $\bm{\rho}= \bm{\phi}(\bm{v})$ is the transfer function of the LMMSE-ESE,  and the function $\bm{v}= \bm{\psi}(\bm{\rho})$ is the transfer function of the DEC, consisting of $K$ separable single-input single-output (SISO) functions  $v_k= \psi_k(\rho_k),  k=1,\ldots,K$. We assume that the channel code of each user is identical to each other, and thus $\psi(\cdot)=\psi_1(\cdot)=\ldots=\psi_K(\cdot)$.
% Note that  $v_k= \psi_k(\rho_k)$ is the transfer function of the decoder of user $k$. 

We first describe the ESE transfer function $\bm{\phi}(\bm{v})$.
Without loss of generality, denote $\bm{\phi}(\bm{v})=[\phi_{1}(\bm{v}),\ldots, \phi_{K}(\bm{v})]^{\text{T}}$.  
From  \eqref{lmmse}, \eqref{ext} and the definition $\rho_k = v_{\text{ext},k}^{-1}$ under \eqref{ext-channel}, $\phi_{k}(\bm{v})$ can be expressed as
\BS \label{rho}
\begin{equation}
	\rho_k = \phi_k(\bm{v}) = \frac{\tau_k}{1-v_k\tau_k},
  \end{equation}
  where	
  \begin{equation}\label{tau}
	\tau_k = \frac{1}{J} \text{tr}\left\{ \bm{A}_k^{\text{H}} \left( \sum_{k'=1}^{K} v_{k'} [\bm{G}_{k'}(\bm{\theta})\bm{W}_{k'}]^2+\sigma^2\bm{I} \right)^{-1}  \bm{A}_k  \right\}.
 \end{equation}\ES
Hence, $\phi_k(\cdot)$ is  a function of $\{\bm{W}_k'\}$ and $\bm{\theta}$.

We now consider the DEC transfer function $\bm{\psi}(\bm{\rho})$. Since each user's data are decoded separately at the DEC, we can express $\bm{\psi}(\bm{\rho})$ as    $\bm{\psi}(\bm{\rho})=[\psi(\rho_1),\ldots,\psi(\rho_K)]^{\text{T}}$, where $\psi(\cdot)$ is a monotone increasing  function of the SINR $\rho_k$.  $\psi(\cdot)$ can be numerically obtained by local Monte Carlo decoding  by taking  \eqref{ext-channel} as the input \cite{precoder}. 

\begin{figure}[htbp]
\centering

\subfigure[]{
\begin{minipage}[t]{0.5\linewidth}
\centering
\includegraphics[width=3.3in]{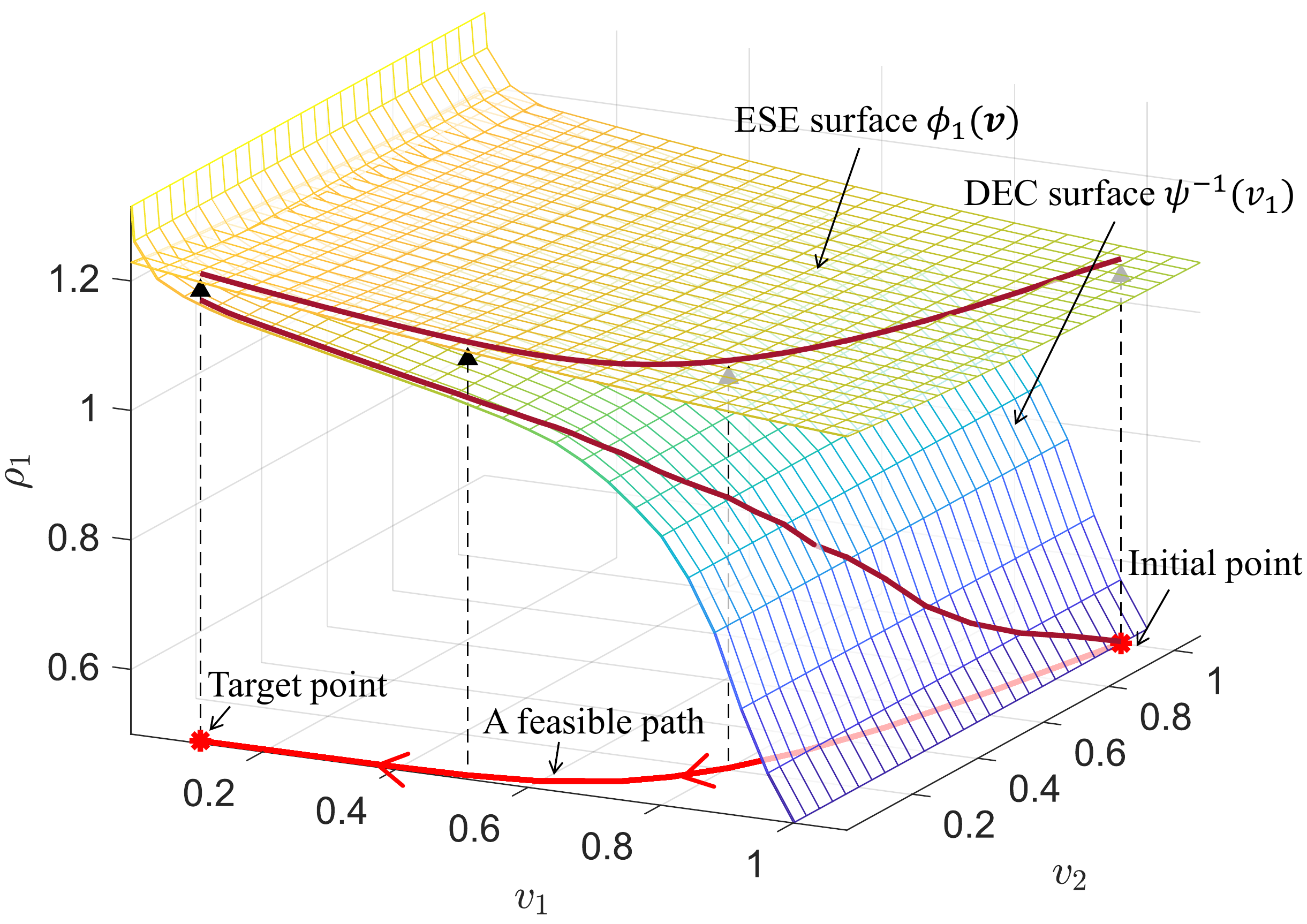}
\end{minipage}
}\subfigure[]{
\begin{minipage}[t]{0.5\linewidth}
\centering
\includegraphics[width=3.3in]{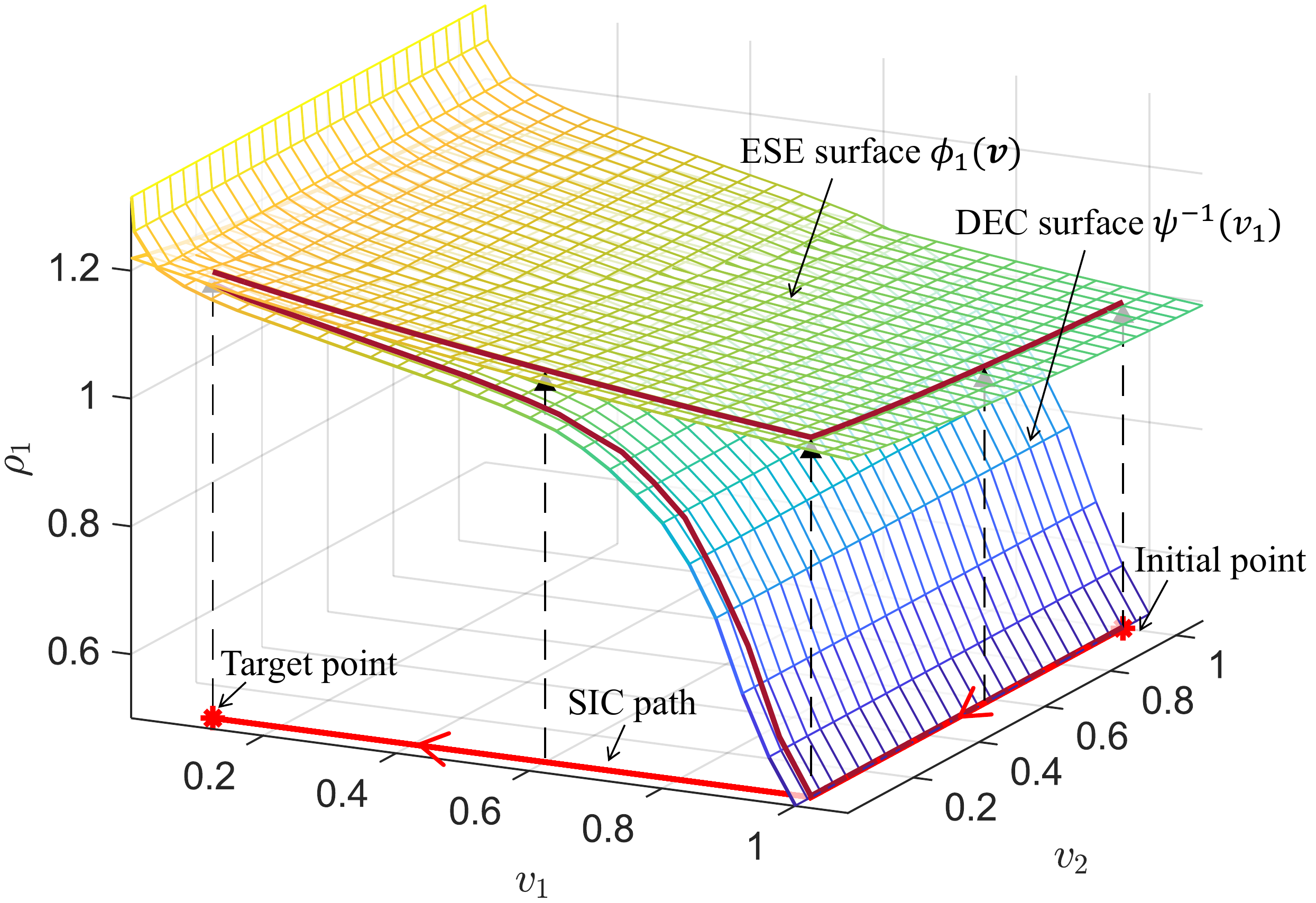}
\end{minipage}
}

\subfigure[]{
\begin{minipage}[t]{0.5\linewidth}
\centering
\includegraphics[width=3.3in]{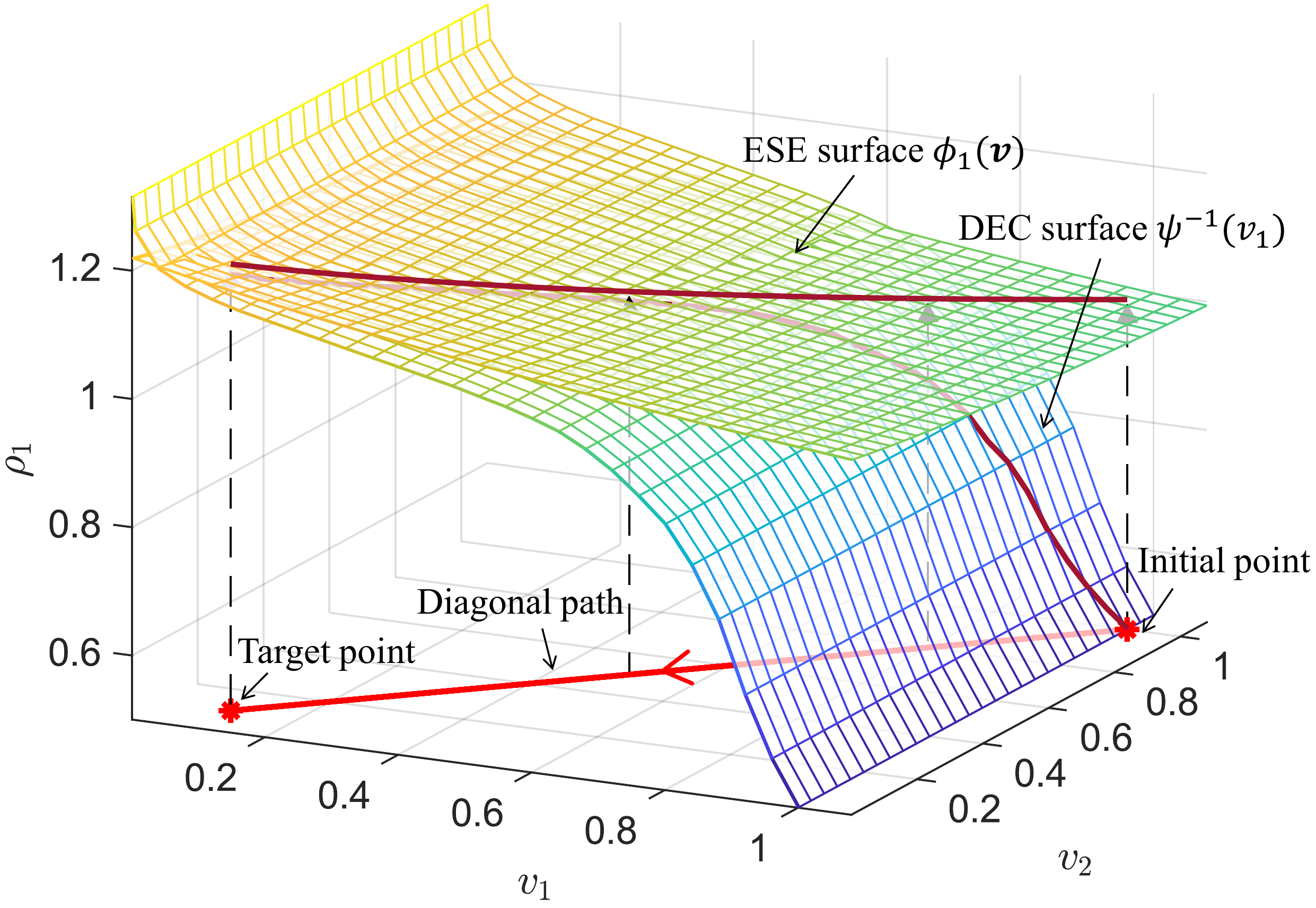}
\end{minipage} \label{se-diag}
}
\centering
\caption{Examples of the path condition \eqref{BER} for user 1 in a 2-user system. (a) A feasible path. (b) The SIC path where the users are decoded one by one, which is a special case of the groupwise SIC in Section \ref{sec-sic}. (c) The diagonal path proposed in \cite{Convergence}, where the input variances of the ESE are set to equal to each other. Thus, the corresponding path is the diagonal of the unit square (or equivalently, a two-dimensional unit cube) formed by the initial point $(1,1)$ and the  point $(0,0)$.}
\label{se}
\end{figure}

Denote by $P_{\text{e},k}$ the  BER  of  user $k$. Note that   $P_{\text{e},k} =\xi(v_k)$ is a  monotone increasing function that maps the input of DEC $v_k$ to the output BER $P_{\text{e},k}$, where $\xi(\cdot)$ can be  obtained by simulations similarly to $\psi(\cdot)$. 
Thus, the output BER at the last iteration is given by $ \xi(v_k^{(T)})$. 
Given a target performance  $P_{\text{tar},k}$, the required $v_k$  can be calculated by $v_{\text{tar},k}=\xi^{-1}(P_{\text{tar},k})$, where $\xi^{-1}(\cdot)$ is the inverse of $\xi(\cdot)$.
To achieve the target BER of each user, we have $\xi(v_k^{(T)}) \le P_{\text{tar},k}, \forall k$, i.e.,
 \begin{align} \label{BER}
     v_{k}^{(T)} \le v_{\text{tar},k}, \; \forall k.
 \end{align}
We say that there exists a feasible path  $\mathcal{L}$ in the $K$-dimension variance space $(v_1,\ldots,v_K)$ from the initial point $(1,\ldots,1)$ to the target point $(v_{\text{tar},1},\ldots,v_{\text{tar},K})$ if \eqref{BER} is met \cite{Convergence}. In other words, if \eqref{BER} is met, there exists a curve $\mathcal{L}$ starting from $(v_1,\ldots,v_K)$ and ending at $(v_{\text{tar},1},\ldots,v_{\text{tar},K})$ such that for each user $k$, the ESE transfer function   $\phi_k(\bm{v})$ is above the  inverse of the DEC transfer function $\psi(\rho_k)$, i.e.,
\begin{align}\label{conver1}
    \phi_k( \bf{v}) > \psi^{-1}(v_k),\; \forall k, \text{for} \; \bf{v}\in {\cal{L}}.
\end{align}
Fig. \ref{se}  illustrates  the path condition \eqref{BER} in a two-user system. We see that there generally exists multiple potential paths that satisfy the path condition.

\subsection{Problem Formulation}\label{p2}
We aim to jointly optimize  $\{\bm{W}_k\}$,  $\bm{\theta}$  and path $\mathcal{L}$ to minimize the total transmit power for the iterative receiver under the constraints of target BER and maximum number of iterations. This problem can be formulated as 
\BS\label{problem1}
\begin{align} 
      \mathcal{P}^{\text{Ite}}: \min_{\{\bm{W}_k\},\bm{\theta}, \mathcal{L}} \quad &\sum_{k=1}^K\sum_{j=1}^{J}|W_k(j,j)|^2 \\
      \text{s.t.}\quad&  \phi_k( \bm{v}) > \psi^{-1}(v_k), \forall k, \; \text{for} \; \bm{v}\in {\cal{L}}, \label{l-const}\\
      & T \le T_{\text{max}}, \label{t-const}\\
      & |\theta_n|=1, \forall n,\label{theta-const4}
\end{align}
\ES
where \eqref{l-const} is the target BER constraint from \eqref{conver1}, \eqref{t-const}  the constraint of the maximal allowed iteration number $T_{\text{max}}$, and  \eqref{theta-const4}  the unit-modulus constraint of the RIS elements. In practice,  $T_{\text{max}}$ is usually set to a small integer to reduce the  computational complexity of the receiver. 

% To guarantee the target BER performance by \eqref{l-const}, the required number of iterations may be unacceptable. In practice,  the processing latency and computational complexity are limited,  and thus we set the maximal number of iterations  to  $T_{\text{max}}$. 
% In addition, for theoretical analysis, we also consider an ideal receiver without complexity limit, and provide an information-theoretic  approach in Appendix A using the existing optimization method \cite{capa},  where the  joint optimization problem is formulated under the constraint of capacity region. The information-theoretic  approach has a huge computational complexity,  which is prohibitive in practical, but it can be seen as an approximate performance limit of the system, and used to evaluated other sub-optimal solutions.

 It is generally difficult to find the globally optimal solution to $\mathcal{P}^{\text{Ite}}$. 
On  one hand, the unit-modulus constraint \eqref{theta-const4} is non-convex, and so is the constraint \eqref{l-const} as seen from \eqref{rho}. Therefore, $\mathcal{P}^{\text{Ite}}$ is a non-convex optimization program  that is difficult to solve. On the other hand,  the optimization result of $\{\bm{W}_k\}$ and $\bm{\theta}$ highly depends on the choice of  the path $\cal{L}$, where $\cal{L}$ is a curve starting from $(v_1,\ldots,v_K)$ and ending at $(v_{\text{tar},1},\ldots,v_{\text{tar},K})$. It is computationally infeasible to search over all the feasible paths in the $K$-dimension space  \cite{achieve-rate2,Convergence}. In \cite{Convergence}, the authors proposed a diagonal path as shown in Fig. \ref{se-diag}, where the input variances of the ESE are set to equal to each other. This is however far from optimal based on our experimental observations.  In the next section,  we  provide an efficient path selection method to obtain an approximate solution to $\mathcal{P}^{\text{Ite}}$.

\section{The Groupwise SIC Approach} \label{sec-sic}
In this section, we present a groupwise-SIC  approach for problem $\mathcal{P}^{\text{Ite}}$. We start with the description of the groupwise SIC approach. Based on  that,  $\mathcal{P}^{\text{Ite}}$ is simplified and then solved by alternately optimizing   precoding and passive beamforming. 
\subsection{Groupwise SIC }
\begin{figure}[t]
    \centering
    \includegraphics[width=4in]{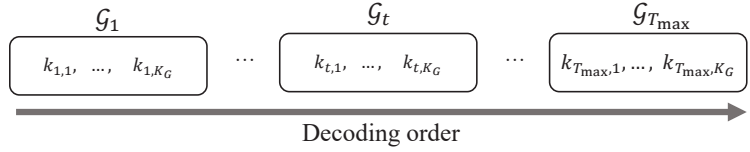}
    \caption{The  groupwise SIC  approach. When decoding $\mathcal{G}_t$ at the $t$th iteration,  only the signals from $\mathcal{G}_{t+1}$ to $\mathcal{G}_{T_{\text{max}}}$ are regraded as  interference, i.e., the interference from $\mathcal{G}_{1}$ to $\mathcal{G}_{t-1}$ are assumed to be perfectly cancelled.}
    \label{group}
\end{figure}

% As mentioned in Section \ref{p2}, it is hard to select an optimal path $\mathcal{L}$ for the iterative receiver. In this section, we provide a heuristic path selecting for Problem $\mathcal{P}^{\text{Ite}}$. 
 We first describe the groupwise SIC approach \cite{gsic}, a.k.a, sequential group detection \cite{sgd1, sgd2}.
 In the groupwise SIC, the users are  divided  into several groups, and  are decoded and cancelled successively in a group-by-group manner.  Following this idea, we divide  the users into $T_{\text{max}}$ groups, and the users in group $t$ are decoded (i.e., to achieve the target BER) at the $t$th iteration. Denote by $\mathcal{K}=\{1,\ldots, K\}$ the total user set, and by $\mathcal{G}_t=\{k_{t,1}, k_{t,2}, \ldots, k_{t,|\mathcal{G}_t|}\}$ the user set of group $t$. As shown in Fig. \ref{group}, the users are  decoded in the order from  $\mathcal{G}_1$ to $\mathcal{G}_{T_{\text{max}}}$, and the interference from the decoded groups is assumed to be perfectly cancelled.
    
The optimal user grouping is generally difficult to determine. We will discuss how to find a sub-optimal user grouping strategy later in Section \ref{sec-grouping}. Here we  focus on the joint optimization  of $\{\bm{W}\}$ and $\bm{\theta}$ for fixed user grouping $\{\mathcal{G}_t\}$.  Given  $ \{\mathcal{G}_t\}$, $\mathcal{P}^{\text{Ite}}$ is reduced to
\begin{subequations}\label{problem3}
\begin{align} 
      \mathcal{P}^{\text{SIC}}: \min_{\{\bm{W}_k\},\bm{\theta}} \quad &\sum_{k=1}^K\sum_{j=1}^{J}|W_k(j,j)|^2 \\
      \text{s.t.}\quad& \phi_k' \ge \rho_{\text{tar},k}, \forall k,  \label{rho-const}\\
      & |\theta_n|=1, \forall n.\label{theta-const5}
\end{align}
where 
\begin{align}
    \rho_{\text{tar},k} & =\psi^{-1}(v_{\text{tar},k}),\\
    \phi_{k}' &=\frac{\tau_k'}{1-\tau_k'}, \label{phi'}\\
    \tau_k' & = \frac{1}{J} \text{tr}\left\{ \bm{A}_k^{\text{H}} \left( \sum_{k'\in \bigcup_{t'=t}^{T_{\text{max}}} \mathcal{G}_{t'}} [\bm{G}_{k'}(\bm{\theta})\bm{W}_{k'}]^2+\sigma^2\bm{I} \right)^{-1}  \bm{A}_k  \right\}. \label{tau'}
\end{align}
\end{subequations}
Eq. \eqref{rho-const} is the user BER constraint,  and \eqref{theta-const5} is the RIS unit-modulus constraint.  Different from \eqref{tau}, for user $k$ in group $\mathcal{G}_t$, the summation in \eqref{tau'} contains the users from group $\mathcal{G}_t$ to $\mathcal{G}_{T_\text{max}}$ since the interference from the previous groups has been cancelled. Due to the non-convexity of \eqref{rho-const} and \eqref{theta-const5}, $\mathcal{P}^{\text{SIC}}$ is  non-convex and is hard to  solve directly. In this paper, to obtain an approximate solution, we  use the alternating optimization (AO) method to find a sub-optimal choice of  $\{\bm{W}_k\}$ and $\bm{\theta}$. 

\subsection{Optimization of $\{\bm{W}_k \}$ Given ${\bm{\theta}}$} \label{SecW}  %
Given  ${\bm{\theta}}$, $\mathcal{P}^{\text{SIC}}$ in \eqref{problem1} is reduced to 
\BS \begin{align}
  \mathcal{P}_{1.1}^{\text{SIC}}: \min_{\{\bm{W}_{k}\}} \quad &\sum_{k=1}^{K}\sum_{j=1}^{J}|W_k(j,j)|^2 \\
  \text{s.t.}\quad& \phi_{k}' \ge \rho_{\text{tar},k}, \; \forall k. \label{phi-const1}
\end{align}\ES
However, $\mathcal{P}_{1.1}^{\text{SIC}}$ is a non-convex optimization problem due to the  fractions and matrix inversions involved in  $\phi_k'$. 
We employ the fractional programming (FP)  \cite{FP} to replace \eqref{phi-const1} with convex constraints. For user $k$ in group $t$, let $\bm{w}_k=[W_k(1,1),\ldots,W_k(j,j)]^{\text{T}}$, $\bm{a}_k=\bm{G}_k\bm{w}_k$ and $\bm{B}_t= \sum_{k' \in  \bigcup_{t'=t}^{T_{\text{max}}} \mathcal{G}_{t'}} [\bm{G}_{k'}\bm{W}_{k'}]^2+\sigma^2\bm{I}$, and we rewrite $\tau_k'$ in \eqref{tau'} as
\begin{align} \label{tau2}
    \tau_{k}' = \frac{1}{J} \bm{a}_{k}^{\text{H}} \bm{B}_t^{-1}\bm{a}_{k}.
\end{align}
With \eqref{phi'} and \eqref{tau2},  constraint \eqref{phi-const1} can be rewritten as 
\begin{align} \label{standard-fp}
    \bm{a}_{k}^{\text{H}} \bm{B}_t^{-1}\bm{a}_{k} \ge \frac{J\rho_{\text{tar},k}}{1+\rho_{\text{tar},k}}, \; \forall k.
\end{align}
Then, by introducing auxiliary variables $\{\bm{y}_k \in \mathbb{C}^{J\times 1}\}$, we can convert $\mathcal{P}_{1.2}^{\text{SIC}}$ to
\begin{subequations} \label{sic-1.3}
  \begin{align}
  \mathcal{P}_{1.2}^{\text{SIC}}: \min_{\{\bm{W}_{k},\bm{y}_k\}} \quad &\sum_{k=1}^{K}\sum_{j=1}^{J}|W_k(j,j)|^2 \\
  \text{s.t.}\quad& 2\text{Re}\{\bm{y}_k^{\text{H}}\bm{a}_k\}- \bm{y}_k^{\text{H}} \bm{B}_t \bm{y}_k \ge \frac{J\rho_{\text{tar},k}}{1+\rho_{\text{tar},k}}, \; \forall k. \label{phi-const3}
\end{align}
\end{subequations}
The equivalence between $\mathcal{P}_{1.1}^{\text{SIC}}$ and $\mathcal{P}_{1.2}^{\text{SIC}}$ is obtained by \cite[Theorem 2]{FP}. Then, we solve $\mathcal{P}_{1.2}^{\text{SIC}} $ by alternately optimizing $\{\bm{y}_k\}$ and $\{\bm{W}_k\}$  as follows.
\begin{enumerate}
    \item Given $\{\bm{W}_k\}$, the optimal solution of $\bm{y}_k$ is  
    \begin{align} 
        \bm{y}_k=\bm{B}_{t}^{-1}\bm{a}_k. \label{y}
    \end{align}
     Note that \eqref{phi-const3} is equivalent to \eqref{standard-fp} by substituting \eqref{y} into \eqref{phi-const3}.
    \item Given $\{\bm{y}_k\}$,   constraint \eqref{phi-const3} is convex \cite{FP}.
    Thus,  $\mathcal{P}_{1.2}^{\text{SIC}}$  is  convex  and  can be solved through convex optimization tools such as CVX \cite{cvx} and interior-point method \cite{ipm}.
\end{enumerate}

\subsection{Optimization of ${\bm{\theta}}$ Given $\{\bm{W}_k\}$ } \label{Sect}
Given $\{\bm{W}_k\}$, $\mathcal{P}^{\text{SIC}}$ in \eqref{problem1} is reduced to
\begin{subequations}
 \begin{align} 
      \mathcal{P}_{2.1}^{\text{SIC}}: \min_{\bm{\theta}} \quad &\sum_{k=1}^K\sum_{j=1}^{J}|W_k(j,j)|^2 \\
      \text{s.t.}\quad& \phi_{k}' \ge \rho_{\text{tar},k}, \; \forall  k, \\
      &|\theta_n|=1, \forall \;n.
\end{align} 
\end{subequations}
In fact, $\mathcal{P}_{2.1}^{\text{SIC}}$ is a feasibility-check problem since the objective function is invariant to the optimization variable $\bm{\theta}$.  To improve the optimization performance, we try to increase the minimum gap between $\phi_k'$ and $\rho_{\text{tar},k}$ by optimizing $\bm{\theta}$,  which  provides a wider feasible region of $\{\bm{W}_k\}$ for $\mathcal{P}_{1.1}^{\text{SIC}}$ at the next iteration and  allows for  further  power reduction. Thus,  we reformulate the problem as
\begin{subequations}
  \begin{align}
  \mathcal{P}_{2.2}^{\text{SIC}}:\max_{\bm{\theta}}& \;\min_{k}\; \phi_{k}'(\bm{\theta})-\rho_{\text{tar},k} \\
  \text{s.t.} & \; |\theta_n|=1, \forall n, \label{theta_2.2}
\end{align}
\end{subequations}
where $\phi_{k}'$ is rewritten as $\phi_{k}'(\bm{\theta})$ to indicate that it is a function of $\bm\theta$. To deal with the non-convex unit-modulus constraint \eqref{theta_2.2}, we replace
$\theta_n$ by $\beta_n$, where $\theta_n = e^{j\beta_n}$ and $\beta_n \in {\mathbb{R}},\forall n$. Let $\bm{\beta} = [\beta_1,\ldots,\beta_N]^{\text{T}}$,  $\mathcal{P}_{2.2}^{\text{SIC}}$  is recast to
\begin{align}
  \mathcal{P}_{2.3}^{\text{SIC}}:\max_{\bm{\beta}\in \mathbb{R}^{N}}& \min_{k} \phi_{k}'(\bm{\beta})-\rho_{\text{tar},k}.
\end{align}
Then, by exploiting  successive convex approximation (SCA),  we  solve a serial of  surrogate problems to obtain a sub-optimal solution to $\mathcal{P}_{2.3}^{\text{SIC}}$ \cite{mm}. 

\begin{lemmaNoParens}
The surrogate function 
  \begin{align} \label{lfunc}
  l_{k}(\bm{\beta}, \bar{\bm{\beta}}) \triangleq -\frac{\alpha_{k}(\bar{\bm{\beta}})\kappa_k}{2}\|\bm{\beta}-\bar{\bm{\beta}}\|^2 +\alpha_{k}(\bar{\bm{\beta}}) \nabla l_{k}'(\bar{\bm{\beta}},\bar{\bm{\beta}})^{\text{T}}(\bm{\beta}-\bar{\bm{\beta}}) +   C_{k}(\bar{\bm{\beta}})
 \end{align}
 satisfies
 \begin{subequations}\label{lowerbound}
  \begin{align} 
      l_{k}(\bar{\bm{\beta}}, \bar{\bm{\beta}}) &=  \phi_{k}'(\bar{\bm{\beta}})-\rho_{\text{tar},k},\\
      l_{k}(\bm{\beta}, \bar{\bm{\beta}}) & \le  \phi_{k}'(\bm{\beta})-\rho_{\text{tar},k},
  \end{align}
\end{subequations}
 where $\kappa_k$ is a constant  no less than the  Lipschitz constant of $\nabla l_{k}(\bm{\beta}, \bar{\bm{\beta}})$,
 \begin{subequations}
 \begin{align}
     \alpha_{k}(\bar{\bm{\beta}})&=\frac{1}{J(1-\tau_{k}'(\bar{\bm{\beta}}))^2},\\
  C_{k}(\bar{\bm{\beta}})&=\phi_{k}'(\bar{\bm{\beta}})- J\alpha_{k}(\bar{\bm{\beta}})\tau_{k}'(\bar{\bm{\beta}}) + \alpha_{k}(\bar{\bm{\beta}})l_{k}'(\bar{\bm{\beta}},\bar{\bm{\beta}}) - \rho_{\text{tar},k},\\
  l_{k}'(\bm{\beta}, \bar{\bm{\beta}})&=\frac{2}{J}\text{Re}\{\bm{y}_{k}(\bar{\bm{\beta}})^{\text{H}}\bm{a}_k(\bm{\beta})\}-\frac{1}{J}\bm{y}_{k}(\bar{\bm{\beta}})^{\text{H}}\bm{B}_k(\bm{\beta})\bm{y}_{k}(\bar{\bm{\beta}}),
 \end{align}
and $\bm{y}_k$ is given by \eqref{y}.
    \end{subequations}
\end{lemmaNoParens}
The proof of Lemma 1 is given in Appendix \ref{sec-proof}. Then, the surrogate problem is given by
\begin{align} \label{suragate-prob}
  \max_{\bm{\beta}}& \min_{k}  l_{k}(\bm{\beta}, \bar{\bm{\beta}}),
\end{align}
where $\bar{\bm{\beta}}$ is the  optimization result of the previous  surrogate problem. As pointed in \cite{mm}, by solving a series of surrogate problems with the surrogate functions satisfying  \eqref{lowerbound}, we can obtain a stationary point of $\mathcal{P}_{2.3}^{\text{SIC}}$.
 Note that $l_{k}(\bm{\beta}, \bar{\bm{\beta}})$   is a concave function, and thus \eqref{suragate-prob} can be easily solved by following \cite{cvx,ipm}.

\subsection{User Grouping} \label{sec-grouping}
 It is computationally involving to find the optimal user grouping with the minimal transmit power. Take exhaustive search for example,  there are totally $(T_{\text{max}})^K$ user groupings, and we need to solve the multiuser MIMO-OFDM power minimization problem for each user grouping. The complexity is  prohibitively high in practice when  $K$ is large. In this paper, we provide a heuristic   low-complexity user grouping strategy based on the  sum rate maximization  with fixed transmit power for each user. 
%  The  user grouping algorithm is described as follows.

We first evenly  assign the  users into $T_{\text{max}}$ groups as follows. 
Following \cite{sgd1}, we order the users based on their channel conditions, where  the users with relatively good channel conditions are decoded first, where the channel condition of each user is characterized by its achievable rate with groupwise SIC. For each group $t$,  denote by $\mathcal{K}_t= \mathcal{K}- \sum_{t'=1}^{t-1}\mathcal{G}_{t'}$ the set consisting of  users that are not in groups $1$ to $t-1$, and  set $\mathcal{K}_1= \mathcal{K}$. Then,  the achievable rate of user $k$ in $\mathcal{K}_t$ is given by \cite{tse2005}
\begin{align}\label{Rk}
    R_{t,k} %&= I(\bm{x}_{k'}; \bm{r}') - I(\bm{x}_{k'}; \bm{r}') \nonumber\\ 
    &= \log\det\left(\bm{I}+\frac{1}{\sigma^2}\sum_{k'\in \mathcal{K}_t }[\bm{G}_{k'}\bm{W}_{k'}]^2\right) - \log\det\left(\bm{I}+\frac{1}{\sigma^2}\sum_{k'\in \mathcal{K}_t, k'\neq k}[\bm{G}_{k'}\bm{W}_{k'}]^2\right).
\end{align}
Clearly, the above rate $R_{t,k}$ of user $k$ is obtained by assuming that the users in the first $t-1$ groups are already decoded and cancelled from the received signal, by following the groupwise SIC strategy.
We order the undecoded users by their achievable rates, and assign  $T_{G}=K/T_{\text{max}}$ users with the largest $R_{t,k}$  into  $\mathcal{G}_t$. We sequentially determine  the initial user grouping  from $\mathcal{G}_1$ to $\mathcal{G}_{T_{\text{max}}}$. The achievable sum rate is given by 
\begin{align}\label{S}
  R_{\text{sum}}=\sum_t\sum_{k \in\mathcal{G}_t} R_{t,k}. 
\end{align}
Then, we adjust the user grouping to further increase the sum rate by reassigning each user to the preceding group or the subsequent group. Specifically, for user $k$ assigned to group $t$, we calculate the sum rate after reassigning it to the preceding group as
\begin{equation}
    R^{\text{pre},k}_{\text{sum}}= \left\{
    \begin{array}{ll}
        R_{\text{sum}}  , & t=1,\\
        \sum\limits_{t'}\sum\limits_{k'\in \mathcal{G}^{\text{pre},k}_{t'}} R_{t',k'} , & 1<t\le T_{\text{max}},
    \end{array}
    \right.
\end{equation}
where $\{\mathcal{G}^{\text{pre},k}_{t'}\}_{t'=1}^{T_{\text{max}}}$ is obtained from $\{\mathcal{G}_{t'}\}_{t'=1}^{T_{\text{max}}}$ by reassigning user $k$ to the preceding group $t-1$ for $1<t\le T_{\text{max}}$.  Similarly, the sum rate after reassigning user $k$  to the subsequent group is given by
\begin{equation}
    S^{\text{sub},k}_{\text{sum}}= \left\{
    \begin{array}{lc}
        R_{\text{sum}}  , & t=T_{\text{max}},\\
        \sum\limits_{t'}\sum\limits_{k'\in \mathcal{G}^{\text{sub},k}_{t'}} R_{t',k'} , & 1\le t<T_{\text{max}},
    \end{array}
    \right.
\end{equation}
where $\{\mathcal{G}^{\text{sub},k}_{t'}\}_{t'=1}^{T_{\text{max}}}$ is obtained from $\{\mathcal{G}_{t'}\}_{t'=1}^{T_{\text{max}}}$ by reassigning user $k$ to the subsequent group $t+1$ for $1\le t<T_{\text{max}}$. The overall user grouping strategy is summarized in Algorithm 1.
% $\{\mathcal{G}_{\text{sub},k,t}\}_{t=1}^{T_{\text{max}}}$ and $S_{\text{sub},k}$ for reassigning user $k$ to the subsequent group, and set  $S_{\text{sub},k} = S$ if user $k$ is in the last group. The largest one of $S$, $S_{\text{pre},k}$ and $S_{\text{pre},k}$ determines  which group user $k$ is  reassigned to.
% If $S_{\text{pre},k}>S$ and $S_{\text{pre},k}>S_{\text{sub},k}$,  we use $\{\mathcal{G}_{\text{pre},k,t}\}_{t=1}^{T_{\text{max}}}$ and $S_{\text{pre},k}$ to update $\{\mathcal{G}_{t}\}_{t=1}^{T_{\text{max}}}$ and $S$, respectively; if $S_{\text{sub},k}>S$ and $S_{\text{sub},k}>S_{\text{pre},k}$,  we use $\{\mathcal{G}_{\text{sub},k,t}\}_{t=1}^{T_{\text{max}}}$ and $S_{\text{sub},k}$ to update $\{\mathcal{G}_{t}\}_{t=1}^{T_{\text{max}}}$ and $S$, respectively; otherwise, $\{\mathcal{G}_{t}\}_{t=1}^{T_{\text{max}}}$ and $S$ are not updated.
We see that the sum rate is non-decreasing by adjusting the user grouping.
The reassigning operation is repeated for all the users until the sum rate $R_{\rm{sum}}$ does not increase anymore. Since the sum rate is non-decreasing, the convergence of the above process  is guaranteed.

\begin{algorithm}[htbp] \label{group-algo}
    \caption{User grouping  algorithm}
    \LinesNumbered 
    \KwIn{ $\boldsymbol{r}'$,  $\boldsymbol{\theta}$,$ \{\bm{G}^{\text{ub}}_k\} $, $   \{\bm{G}_{n}^{\text{rb}}\}$, $\{ \bm{G}_{k,n}^{\text{ur}} \}$, $\{\boldsymbol{W}_k\}$, $T_{\rm{max}}$.}
    \For{$t=1, \ldots, T_{\rm{{max}}}$}{
                  Order the unassigned users by \eqref{Rk}, and  assign  $T_{G}$ users with the largest $R_{t,k}$  into  $\mathcal{G}_t$\;
        }
    Calculate the sum rate $R_{\rm{sum}}$ by \eqref{S}\;
    \While{$R_{\rm{sum}}$ is  increasing}{
                \For{$k=1, \ldots, K$}{
                  Calculate $R^{{\rm{pre}},k}_{\rm{sum}}$ and $R^{{\rm{sub}},k}_{\rm{sum}}$\;
                    \If{$R_{\rm{sum}}^{{\rm{pre}},k}>R_{\rm{sum}}$ and $R^{{\rm{pre}},k}_{\rm{sum}}\ge R^{{\rm{sub}},k}_{\rm{sum}}$}
                    { $\mathcal{G}_{t} = \mathcal{G}^{{\rm{pre}},k}_{t}, \forall t$ and $R_{\rm{sum}} = R^{{\rm{pre}},k}_{\rm{sum}}$\;}
                    \If{$R^{{\rm{sub}},k}_{\rm{sum}}>R_{\rm{sum}}$ and $R_{\rm{sum}}^{{\rm{sub}},k}>R_{\rm{sum}}^{{\rm{pre}},k}$}
                    {$\mathcal{G}_{t} = \mathcal{G}^{{\rm{sub}},k}_{t}, \forall t$ and $R_{\rm{sum}} = R_{\rm{sum}}^{{\rm{sub}},k}$\;}
                    }
                }
            
            \KwOut{$\{\mathcal{G}_t\}$.}
        \end{algorithm}

\subsection{Convergence and Complexity}
The overall groupwise SIC approach  is summarized in Algorithm \ref{sic-algo}. 
The convergence of the algorithm is guaranteed, since the objective function is non-increasing during the iteration of AO and is  also lower-bounded by zero.   The algorithm stops when the change of the objective function is less than a predetermined threshold. 
Using the interior-point method \cite{ipm-comp}  for  $\mathcal{P}_{1.2}^{\text{SIC}}$ and  $\mathcal{P}_{2.3}^{\text{SIC}}$, the complexity of one iteration of Algorithm \ref{sic-algo} is $\mathcal{O}((K(J+1))^{3.5}+T_1 N^{3.5})$, where  $T_1$ is the number of successive convex approximation iterations. 
Compared with the information-theoretic  approach in Appendix \ref{sec-info}, whose  complexity is ${\cal{O}}((KJ+2^K)^{3.5}+(2^{K})^{3.5}N)$,  the groupwise SIC  algorithm has a much lower complexity as the number of users increases. 
% For example, consider the setting with $K=12$, $N=200$, $J=16$,  and $T_1=500$, the complexity of the two approaches is about $\mathcal{O}(10^{14})$ and $\mathcal{O}(10^{10})$, respectively.

\begin{algorithm}[htbp] \label{sic-algo}
    \caption{Groupwise SIC optimization}
    \LinesNumbered 
    \KwIn{$\boldsymbol{r}'$,  $ \{\bm{G}^{\text{ub}}_k\} $, $   \{\bm{G}_{n}^{\text{rb}}\}$, $\{ \bm{G}_{k,n}^{\text{ur}} \}$.}
    
    Randomly generate an independent realization of $\bm{\theta}$\; 
    Obtain the user grouping $\{\mathcal{G}_t\}$ by  Algorithm \ref{group-algo}\;
    \While{the stopping criterion is not met}
    {
      
        Update $\{\bm{W}_k\}$ by solving $\mathcal{P}_{1.2}^{\text{SIC}}$ through convex optimization tools given $\bm{\theta}$\;

        Obtain $\bm{\beta}$ by solving $\mathcal{P}_{2.3}^{\text{SIC}}$   given $\{\bm{W}_k\}$, and then update $\bm{\theta}$ by $\bm{\theta} = e^{j \bm{\beta} }$\;
        
    }  
    \KwOut{$\{\bm{W}_k\}$ and $\bm{\theta}$.}
\end{algorithm}

\section{Numerical Results}\label{sec-simu}

In this section, we evaluate the proposed scheme with simulations.  
The metrics  are the average BER of all the users, i.e., $\frac{1}{K}\sum_k P_{\text{e},k}$, and the average transmit power defined as $P=\frac{1}{JK}\sum_k\sum_j|W(j,j)|^2)$. Consider a three-dimensional coordinate system, where the BS is located at $(0,0,10)$,  and the receiving antennas are  a uniform linear array located on the $x$-axis with half-wavelength antenna spacing. The RIS, located at $(50,50,10)$,  is a uniform planar array  parallel to the $x-z$ plane with half-wavelength element spacing. The locations of the users are randomly and uniformly distributed in the horizontal rectangular area formed by the  point $(60,0, 1.5)$ and the point $(110, 50, 1.5)$.  Following \cite{yanggang}, we assume that the user-RIS and RIS-BS channels are both Rician fading distributed, where the large-scale pathlosses are $10^{-3}d^{-2}$ and $10^{-3}d^{-2.5}$, respectively, with $d$ being the distance, and the Rician factors  are both 10.  The user-BS channels  are Rayleigh fading with the  large-scale pathloss
$10^{-3}d^{-4}$. The numbers of the delay taps of these channels are  $L_{\text{ur}}=2$,  $L_{\text{rb}}=5$ and $L_{\text{rb}}=6$, respectively.   We assume that all the users use the  same  channel code. The target BER  is set as $10^{-4}$, and the LDPC code in 3GPP TR 38.212 \cite{TR38212} for QPSK modulation, code rate $=1/2$ and information length $=2112$ is used for the channel code.  Since each codeword of a user is transmitted over a single OFDM symbol, the number of subcarriers is $J=2112$. In practice, for complexity consideration, it is not necessary to allocate a different power for every subcarrier. Thus, we downsample the frequency channel to $J'$ subcarriers in power optimization, i.e., $\bm{G}_{k}(\bm{\theta}) \in \mathbb{C}^{JM\times 1}$ to $\bm{G}_{k}'(\bm{\theta}) \in \mathbb{C}^{J'M\times 1}$. Then, every $J/J'$ subcarriers share the same power. We set $\sigma^2=-105$ dBm, and all the results are averaged more than 200 independent channel realizations. With the transmitter and iterative receiver in Section \ref{sec-system}, the proposed groupwise-SIC  approach is compared with the following four baseline approaches:
\begin{enumerate}
    \item \textbf{No-RIS approach}: Obtain the optimized $\{\bm{W}_k\}$  by solving $\mathcal{P}_{1.1}^{\text{SIC}}$ in \ref{SecW} with given $\bm{G}_k(\bm{\theta}) = \bm{G}_k^{\text{ub}}$ and $T_{\text{max}}=1$.
    \item \textbf{Random-phases approach}: Obtain the optimized $\{\bm{W}_k\}$  by solving $\mathcal{P}_{1.1}^{\text{SIC}}$ in \ref{SecW} with a randomly generated $\bm{\theta}$ and $T_{\text{max}}=1$.
    \item \textbf{Information-theoretic approach}: The optimization approach in Appendix \ref{sec-info}, where the joint precoding and passive beamforming optimization problem  under the constraint of capacity region is formulated. This optimization approach is inspired by the single-user optimization algorithm in \cite{capa}.
    \item \textbf{Diagonal-path approach}: Using the diagonal path  shown in Fig. \ref{se-diag} and removing the constraint of iteration number, Problem $\mathcal{P}^{\text{Ite}}$ degenerates to the form similar to the single user optimization problem in \cite{single-sys}, and can be solved using the FP and MM derived in this paper. However,  the computational complexity of the diagonal-path approach is much higher than the groupwise SIC approach, since the number of constraints of the former one is much more due to the discretization of the path $\mathcal{L}$ \cite{single-sys,Convergence}. Due to space limitations, the details of the  diagonal-path approach are omitted.
\end{enumerate}

\begin{figure}[t]
    \centering
    \subfigure[$T_{\text{max}}=1$.]{
\begin{minipage}[t]{0.5\linewidth}
\centering
\includegraphics[width=2.5in]{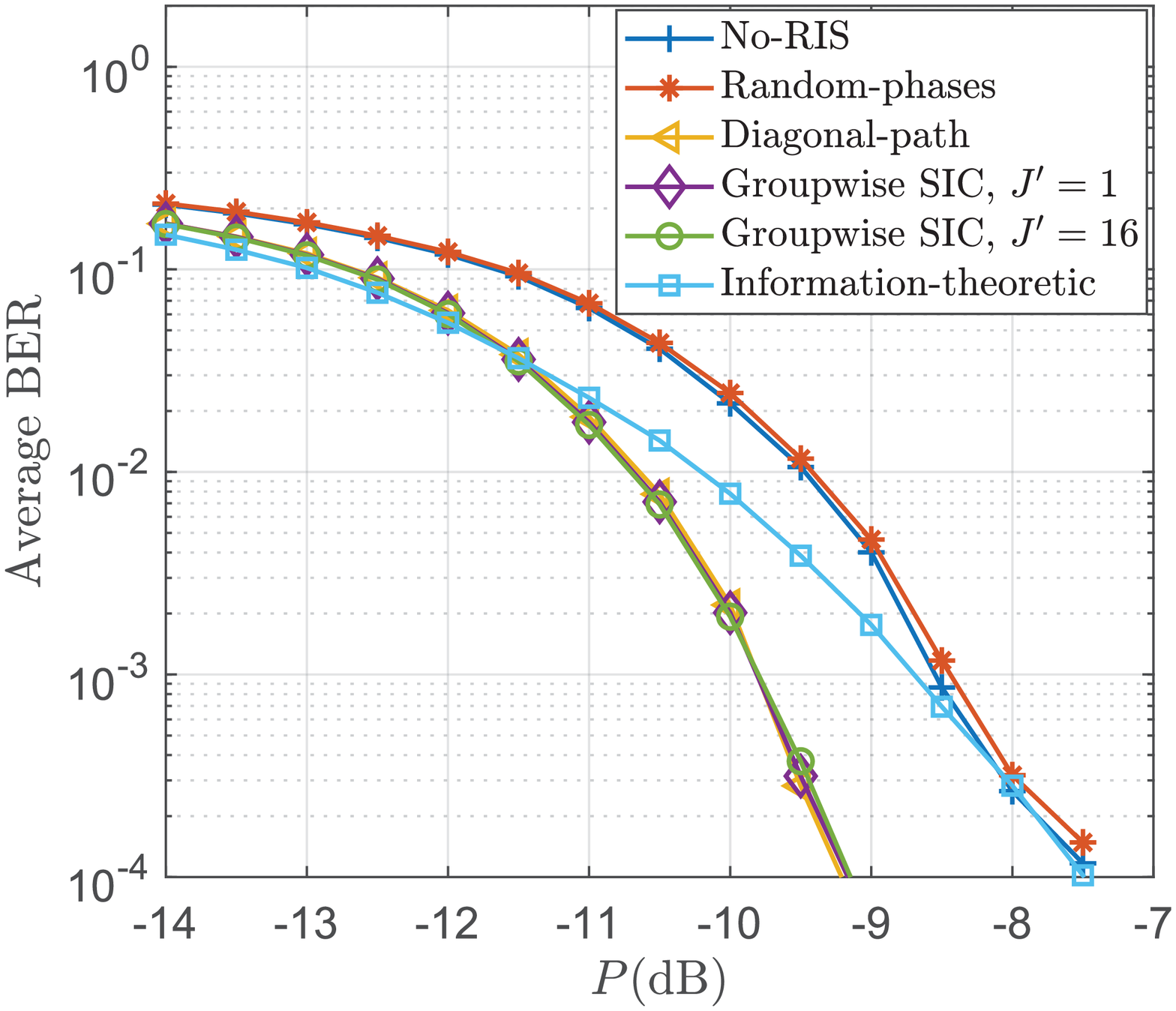}
%\caption{fig1}
\end{minipage}%
\label{sim-all-1}
}%
\subfigure[$T_{\text{max}}=4$.]{
\begin{minipage}[t]{0.5\linewidth}
\centering
\includegraphics[width=2.5in]{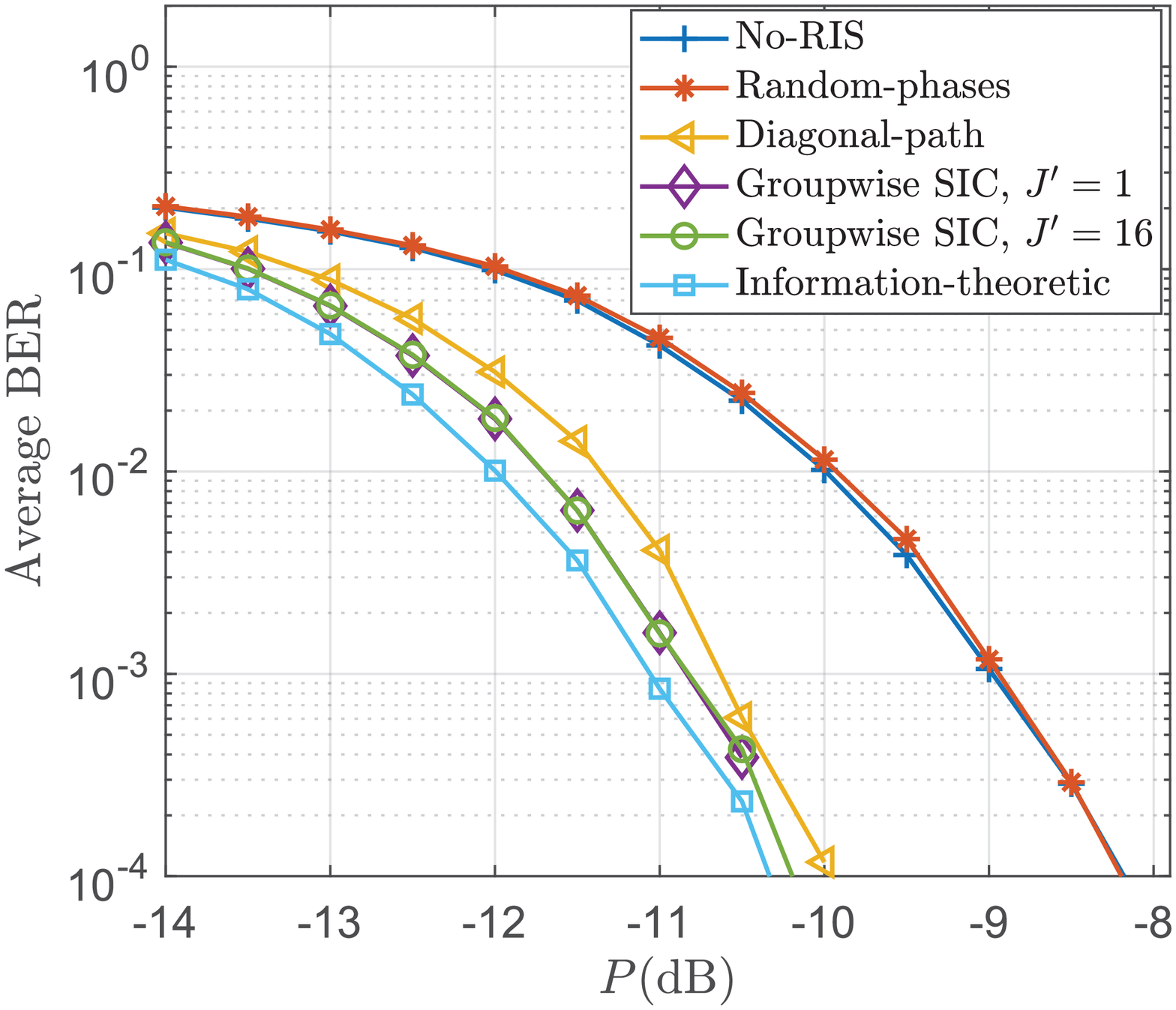}
%\caption{fig2}
\end{minipage}%
\label{sim-all-2}
}%
    \caption{Performance comparison between the proposed approach and baseline approaches. Settings: $K=4, M=8, N=80$.}
    \label{sim-all}
\end{figure}
We first evaluate the information-theoretic and other  approaches  in scenarios with a relatively small number of users (e.g., 4 users). In the following, the optimization approaches use $J'=16$ unless specified otherwise, which means that  every $2112/16=132$ subcarriers shares the same power. As shown in Fig. \ref{sim-all-1}, we compare the proposed approaches and baseline approaches in a 4-user system with $T_\text{max}=1$. First, we see that the groupwise SIC  approaches and the diagonal-path approach   significantly outperform the other approaches, with about 1.5 dB power gain at the target BER.  Second, the groupwise SIC  approaches with $J'=1$ and $J'=16$ have almost the same performance. % which is also verified in Fig. \ref{sim-ant}. 
Fig. \ref{sim-all-1} shows the performance comparison between the approaches with $T_\text{max}=4$. The information-theoretic approach has a better performance than the other two approaches. The reason is that the iterative receiver with a higher complexity (or more iterations) behaves more like an ideal receiver, and so the information-theoretic approach works better. In addition, the diagonal-path approach has a performance gap less than 0.5 dB compared with other two optimization approaches.

\begin{figure}[htpb]
    \centering
    \includegraphics[width=2.8in]{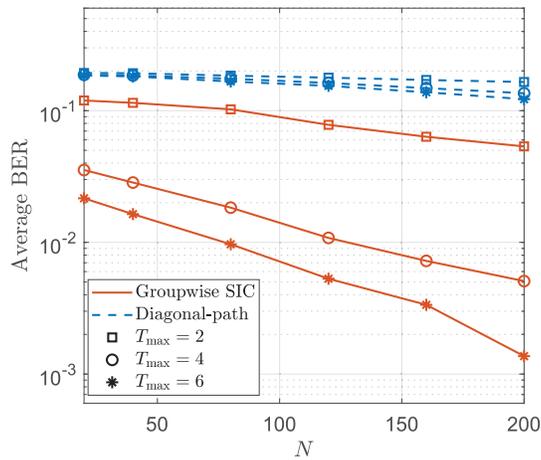}
    \caption{Performance comparison under different numbers of RIS elements. Settings: $K=12$ and $M=8$.}
    \label{sim-n}
\end{figure}

\begin{figure}[htpb]
    \centering
    \includegraphics[width=2.8in]{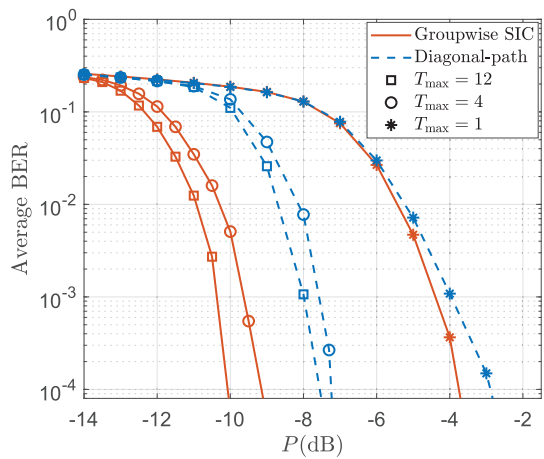}
    \caption{Performance comparison under different maximum iteration numbers in a 12-user system. Settings:  $M=8$, and $N=200$.}
    \label{sim-12user}
\end{figure}

We now evaluate the  groupwise SIC and diagonal-path approaches in the setting with  more users (e.g., 4 to 16 users), where $J'=1$ for all the approaches. 
Fig. \ref{sim-n} shows the performance comparison under different  numbers of RIS elements $N$ in a 12-user system. The groupwise SIC  approach has a significant performance gain compared with the diagonal approach, no matter the maximum number of iterations is 2, 4 or 6. In addition, the performance of the groupwise SIC  approach  improves as the increasing of the iteration number $T_{\text{max}}$, which provides a clear demonstration of the performance-complexity trade-off of our considered iterative detection scheme. 

The performances under different   $T_{\text{max}}$ in a 12-user system are shown Fig. \ref{sim-12user}. First, we see that  the groupwise SIC approach always outperforms the diagonal-path approach. As the number of iterations increases to 12, more than 2 dB power gain can be obtained. Second, with 4 iterations, a large portion of the performance gain  can be obtained, which demonstrates the advantage of the groupwise SIC  approach for a low-complexity receiver.

\begin{figure}[t]
    \centering
    \includegraphics[width=2.8in]{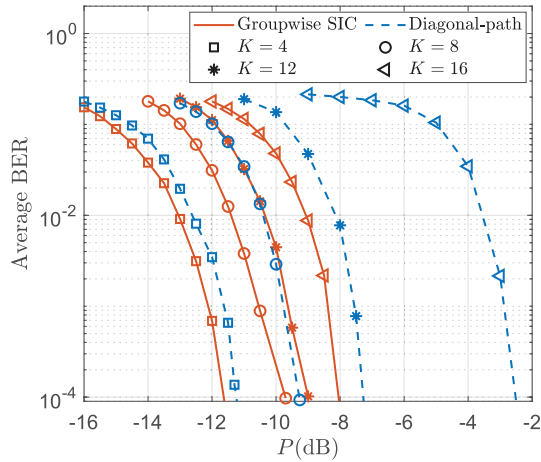}
    \caption{Performance comparison  with a varying number of users.  Settings: $M=8, N=200$, and $T_{\text{max}}=4$.}
    \label{sim-24user}
\end{figure}

Fig. \ref{sim-24user} shows the  performance comparison under  a varying number of users.  The groupwise SIC  approach shows a significant performance gain compared with the diagonal-path approach, especially when the number of users is large. The performance gain is less than 0.5 dB when $K=4$, and increases to about 5 dB when $K=16$. Therefore, it can be concluded that the groupwise SIC approach has  advantages in both  performance and  computational complexity compared with its counterparts, especially when the number of iterations is relatively small.

\section{Conclusion} \label{sec-con}
In this paper, we studied the RIS-aided  multiuser MIMO-OFDM system with  a specific  iterative receiver. 
We formulated the joint optimization problem for the iterative receiver under the constraints of user BER and maximal iteration number. We  proposed the low-complexity groupwise SIC  approach and converted the problem to two  sub-problems of precoding and passive beamforming. For precoding,  we apply the FP to deal with the non-convex constraints with matrix inversion and fractions. For passive beamforming, we redesign the feasibility-check problem, and resort to the SCA to deal with the unit-modulus constraints of RIS. We also provided a heuristic  and low-complexity user grouping approach. We show that the proposed groupwise SIC  approach has a much lower complexity than the information-theoretic  approach.
Numerical simulations showed that the groupwise SIC  approach outperforms the information-theoretic approach and the diagonal-path approach,  especially when the iteration number of the receiver is limited to a relatively small value. 

\begin{appendices}

\section{Information-Theoretic   Approach} \label{sec-info}
% In Appendix A, for the ideal receiver, the  joint precoding and passive beamforming  optimization problem for the RIS-aided MIMO-NOMA system is formulated from an information-theoretic perspective. Due to the unit-modulus constraints of $\bm{\theta}$, the optimization is non-convex and difficult to solve directly. Thus, we apply alternating optimization (AO)  to obtain an approximate solution to the problem by  optimizing the variables alternately. 

\subsection{Problem Formulation}
Denote by $\mathcal{K}=\{1,\ldots, K\}$ the total user set. Following \cite{tse2005} and \cite{info},   the capacity region of the considered multiuser MIMO-OFDM transmission  can be  expressed as
\begin{align}\label{capa1}
    \sum_{k \in \mathcal{K}_{u}}QR_{k} & \le  \frac{1}{J+L_{\text{cp}}} \log \det \left(\bm{I}+\frac{1}{\sigma^2}\sum_{k \in \mathcal{K}_{u}}\left[\bm{G}_k(\bm{\theta})\bm{W}_{k}\right]^2\right), \;  \mathcal{K}_u \subseteq \mathcal{K},
\end{align}
where $QR_k$  is the  transmission rate per channel use.
Then, the information-theoretic  optimization problem is formulated by 
\begin{subequations}
  \begin{align}
    \mathcal{P}^{\text{Info}}: \min_{\{\bm{W}_{k}\}, \bm{\theta}} \quad &\sum_{k=1}^K\sum_{j=1}^{J}|W_k(j,j)|^2 \label{obj} \\
    \text{s.t.} \quad &    \sum_{k \in \mathcal{K}_{u}}\left(J+L_{\text{cp}}\right)QR_{k} \le   \log \det \left(\bm{I}+\frac{1}{\sigma^2}\sum_{k \in \mathcal{K}_{u}} \left[\bm{G}_k(\bm{\theta})\bm{W}_{k}\right]^2\right),  \mathcal{K}_u \subseteq \mathcal{K}, \label{capa-const}\\
    & |\theta_n|=1, \forall n. \label{theta-const1}
\end{align}
\end{subequations}
where \eqref{obj} is  the total transmit power of all users, \eqref{capa-const} is the capacity region constraint, and  \eqref{theta-const1} is the unit-modulus constraint of the RIS's elements.
% The optimization problem $\mathcal{P}^{\text{Info}}$ is hard to solve directly because the unit-modulus constraint of $\theta_n$ \eqref{theta-const1} is non-convex and the variables $\{\bm{W}_k\}$ as well as $\bm{\theta}$ are coupled in \eqref{capa-const}. 
As inspired by \cite{capa}, we resort to the AO  method to obtain an approximate solution to $\mathcal{P}^{\text{Info}}$ by optimizing $\{\bm{W}_k\}$ and each $\theta_n$ alternately, as  described in the following subsections.

\subsection{Optimization of  $\{\bm{W}_k\}$ Given $\bm{\theta}$} \label{information-w}
Given $\bm{\theta}$, $\mathcal{P}^{\text{Info}}$ is  reduced to
\begin{subequations}
  \begin{align}
  \mathcal{P}_{1}^{\text{Info}}:  \min_{\{\bm{W}_{k}\}} \quad &\sum_{k=1}^K\sum_{j=1}^{J}|W_k(j,j)|^2  \\
  \text{s.t.} \quad & \sum_{k \in \mathcal{K}_{u}}\left(J+L_{\text{cp}}\right)QR_{k} \le   \log _2 \det \left(\bm{I}+\frac{1}{\sigma^2}\sum_{k \in \mathcal{K}_{u}} \left[\bm{G}_k(\bm{\theta})\bm{W}_{k}\right]^2\right), \mathcal{K}_u \subseteq \mathcal{K}.
\end{align}
\end{subequations}
% Note that $\log (\cdot)$ is a concave function and $ \det (\cdot)$ is  concave of Hermite matrices \cite{tse2005}. 
Since $\log\det(\cdot)$ is a concave function of $\{\bm{W}_k\}$ \cite{tse2005},   constraint \eqref{capa-const} is convex. Therefore,  $\mathcal{P}_{1}^{\text{Info}}$  can be solved by the existing convex  optimization tools \cite{ipm,cvx}. 

\subsection{Optimization of $\theta_n$ Given $\{\bm{W}_k\}$ and $\{\theta_{n'}, n' \neq n\}$} \label{information-theta}
 Given $\{\bm{W}_k\}$ and $\{\theta_{n'}, n' \neq n\}$,  $\mathcal{P}^{\text{Info}}$ is  reduced to
 \begin{subequations}
    \begin{align}
  \mathcal{P}_{2.1}^{\text{Info}}:  \min_{\theta_n} \quad &\sum_{k=1}^K\sum_{j=1}^{J}|W_k(j,j)|^2  \label{obj2.1}\\
  \text{s.t.} \quad & \sum_{k \in \mathcal{K}_{u}}\left(J+L_{\text{cp}}\right)QR_{k} \le   \log _2 \det \left(\bm{I}+\frac{1}{\sigma^2}\sum_{k \in \mathcal{K}_{u}} \left[\bm{G}_k(\bm{\theta})\bm{W}_{k}\right]^2\right),  \mathcal{K}_u \subseteq \mathcal{K}, \\ 
  &  |\theta_n|= 1. \label{theta-const2}
\end{align}
 \end{subequations}
Note that $\mathcal{P}_{2.1}^{\text{Info}}$ is a feasibility-check problem.
Similar to $\mathcal{P}_{2.1}^{\text{SIC}}$ in Section \ref{Sect}, by introducing an auxiliary variable $\Delta R$, we reformulate $\mathcal{P}_{2.1}^{\text{Info}}$ as
\begin{subequations}
 \begin{align}
  \mathcal{P}_{2.2}^{\text{Info}}: \max_{\theta_n, \Delta R} \quad & \Delta R \\
  \text{s.t.} \quad &  \sum_{k\in \mathcal{K}_u}Q(J+L_{\text{cp}})(R_k+\Delta R) \le \log _2 \det \left(\bm{I}+\frac{1}{\sigma^2}\sum_{k\in \mathcal{K}_u}\left[\bm{G}_k(\bm{\theta})\bm{W}_{k}\right]^2 \right),  \mathcal{K}_u \subseteq \mathcal{K},\label{deltar-const}\\
  &   |\theta_n|= 1. \label{2.2theta}
\end{align}
\end{subequations}
% Intuitively, , and $\mathcal{P}_{1.3}$  However, $\mathcal{P}_{1.2}$  is still non-convex for $\theta_n$ because of constraint \eqref{deltar-const}. 
To solve this  problem,  we rewrite $\bm{G}_k(\bm{\theta})$ as
 \begin{align}\label{G_new}
     \bm{G}_k(\bm{\theta}) = \bm{G}'_{n,k}  +  \theta_n\bm{G}_{n}^{\text{rb}} \bm{G}_{k,n}^{\text{ur}},
 \end{align}
 where $\bm{G}'_{n,k} = \bm{G}^{\text{ub}}_k  +  \sum_{n'\neq n}\theta_{n'} \bm{G}_{n'}^{\text{rb}} \bm{G}_{k,n'}^{\text{ur}}$. As such,  the  right-hand side of \eqref{deltar-const} can be rewritten as \cite{capa} 
%  &\triangleq \log _2 \det \Bigg(\bm{I}+ \frac{1}{\sigma^2}\sum_{k\in \mathcal{K}_u}  \left[\bm{G}_k(\bm{\theta})\bm{W}_{k}\right]^2\Bigg)\nonumber\\
%     &= \log _2 \det \Bigg(\bm{I}+ \frac{1}{\sigma^2}\sum_{k\in \mathcal{K}_u}  \left[\bm{G}'_{n,k}  +  \theta_n \bm{G}_{n}^{\text{rb}} \bm{G}_{k,n}^{\text{ur}}\right) \bm{W}_k]^2\Bigg)\nonumber\\
%     & = \log _2 \det \Bigg(\!\!\bm{I}\!+\!\frac{1}{\sigma^2}\!\!\sum_{k\in \mathcal{K}_u} \!\!
%     \Big(\!2\text{Re}\{\theta_n\bm{G}^{\text{rb}}_n\bm{G}^{\text{ur}}_{k,n}\bm{W}_k(\bm{G}'_{n,k}\bm{W}_k)^{\text{H}}\} \!+\![\bm{W}_k\bm{G}'_{n,k}]^2 \!+\! \theta_n\theta_n^{\text{H}}[\bm{G}^{\text{rb}}_n\bm{G}^{\text{ur}}_{k,n}\bm{W}_k]^2 \Big)\!\!\!\Bigg)\nonumber \\
\begin{align}\label{f}
    &C_{u}(\theta_n)  = \log _2 \det \Bigg(\!\!\bm{I}\!+\! \frac{1}{\sigma^2}\!\!\sum_{k\in \mathcal{K}_u} \!\!
    \Big(\!2\text{Re}\{\theta_n\bm{G}^{\text{rb}}_n\bm{G}^{\text{ur}}_{k,n}\bm{W}_k(\bm{W}_k\bm{G}'_{n,k})^{\text{H}}\}  \!+\! [\bm{W}_k\bm{G}'_{n,k}]^2+  [\bm{G}^{\text{rb}}_n\bm{G}^{\text{ur}}_{k,n}\bm{W}_k]^2 \Big)\!\!\!\Bigg),
\end{align} 
% where $(a)$ holds due to $|\theta_n|=1$. 
% Given $\{\bm{W}_k\}$ and $\{\theta_{n'}, n'\neq n\}$, \eqref{f} is a convex function over $\theta_n$. However, $\mathcal{P}_{2.2}^{\text{SIC}}$ is still non-convex because of constraint \eqref{2.2theta}.  
Then we  apply the convex relaxation technique by  relaxing $|\theta_n|=1$ to $|\theta_n|\le 1$. Thus, $\mathcal{P}_{2.2}^{\text{Info}}$ is converted to 
\begin{subequations}
 \begin{align}
  \mathcal{P}_{2.3}^{\text{Info}}: \max_{\theta_n, \Delta R} \quad & \Delta R \label{deltaR}\\
  \text{s.t.} \quad &   \sum_{k\in \mathcal{K}_u}Q(J+L_{\text{cp}})(R_k + \Delta R ) \le C_{u}(\theta_n),  \mathcal{K}_u \subseteq \mathcal{K},\label{c-const}\\
  &|\theta_n|\le1. \label{theta-const3}
\end{align} 
\end{subequations}
Since \eqref{deltaR}-\eqref{theta-const3} are all convex, $\mathcal{P}_{2.3}^{\text{Info}}$ is  a convex optimization problem that can be solved by existing convex optimization tools \cite{cvx,ipm}. During the iteration of AO, if the obtained $\theta_n$ for $\mathcal{P}_{2.3}^{\text{Info}}$ does not  satisfy the constraint in \eqref{deltar-const}, we normalize  $\theta_n$ by $\theta_n/|\theta_n|$   and then optimize $\{\bm{W}_k\}$ based on the normalized $\bm{\theta}$. As stated in \cite{capa}, due to the above relaxation and normalization of $\bm{\theta}_n$, the convergence of the AO is not guaranteed.
 
\subsection{Complexity} %Convergence and
% \begin{algorithm}[hbt] \label{capa-algo}
%     \caption{ Information-theoretic  optimization algorithm}
%     \LinesNumbered 
%     \KwIn{ $\boldsymbol{r}'$, $ \{\bm{G}^{\text{ub}}_k\} $, $   \{\bm{G}_{n}^{\text{rb}}\}$, $\{ \bm{G}_{k,n}^{\text{ur}} \}$.}
    
%     Randomly generate an independent realization of $\bm{\theta}$\; 
%     \While{the stopping criterion is not met}
%     {
%         Obtain $\{\bm{W}_k\}$ by solving $\mathcal{P}_{1}^{\text{Info}}$ given $\bm{\theta}$\;
%         \For{$n = 1, \ldots,\;N$} {
%             Obtain ${\theta_n}$ by solving $\mathcal{P}_{2.3}^{\text{Info}}$ given $\{\bm{W}_k\}$ and $\{\theta_{n'}, n'\neq n\}$\;
%             If $|\theta_n|<1$, $\theta_n = \theta_n/|\theta_n|$;
%         }
%     }  
%     \KwOut{$\{\bm{W}_k\}$ and $\bm{\theta}$.}
% \end{algorithm}
% The information-theoretic approach for  $\mathcal{P}^{\text{Info}}$ is summarized in Algorithm \ref{capa-algo}. Due to the normalization of $\bm{\theta}$, the convergence of  Algorithm \ref{capa-algo} is not guaranteed. But, based on the simulations in \cite{capa} and this paper, this algorithm  can still obtain a  good performance. In addition, t
 The numbers of constraints in \eqref{capa-const} and \eqref{c-const} are both about $2^K$. With the interior-point method \cite{ipm-comp}, the complexity of one iteration  is ${\cal{O}}((KJ+2^K)^{3.5}+(2^{K})^{3.5}N)$, where the complexity of solving the $\mathcal{P}_{1}^{\text{Info}}$ and $\mathcal{P}_{2.3}^{\text{Info}}$ is  ${\cal{O}}((KJ+2^K)^{3.5})$ and ${\cal{O}}((2^{K})^{3.5})$ , respectively.  
%  Compared with the groupwise SIC  approach  whose  complexity is $\mathcal{O}((K(J+1))^{3.5}+T_1 N^{3.5})$,  the information-theoretic approach has a much higher complexity as the number of users increases. For example, consider the setting with $K=12$, $N=200$,$J=16$,  and $T_1=500$ , the complexity of the two approachs is about $\mathcal{O}(10^{14})$ and $\mathcal{O}(10^{10})$, respectively.
% For example, each sub-problem will include $2^{10}-1=1023$ constraints in a 10-user system. 

% Algorithm \ref{capa-algo} provides an information-theoretic  algorithm, which gives the theoretic limit of the RIS-aided MIMO-NOMA system.  However, it is still unclear how to design a  low-complexity receiver to achieve the information-theoretic limit. For example, as will be shown in Section \ref{sec-simu}, directly applying $\{\bm{W}_k\}$ and $\bm{\theta}$ of Algorithm \ref{capa-algo} in a specific low-complexity receiver may cause performance degradation. Therefore, it is desirable to jointly optimize the precoding and the passive beamforming for a specific and  low-complexity receiver. In next section, for the iterative receiver in Section \ref{sec-system}, we reformulate  the joint optimization problem under practical constraints.

\section{Proof of Lemma 1} \label{sec-proof}
We construct $l_{k}(\bm{\beta}, \bar{\bm{\beta}})$ by using the following three steps.
\begin{enumerate}
    \item Lower bound $l_{1,k}(\bm{\beta}, \bar{\bm{\beta}})$ of $\phi_{k}'$: Note that $\phi_{k}'=\frac{\tau_{k}'}{1-\tau_{k}'}$ is a convex function of $\tau_{k}'$. Thus, the first order Taylor expansion of $\phi_{k}'(\tau_{k}'(\bm{\beta}))$ at $\tau_{k}'(\bar{\bm{\beta}})$ satisfies
    \begin{align}\label{l1}
  l_{1,k}(\bm{\beta}, \bar{\bm{\beta}})= \phi_{k}'(\bar{\bm{\beta}})+ \frac{\tau_{k}'(\bm{\beta})-\tau_{k}'(\bar{\bm{\beta}})}{(1-\tau_{k}'(\bar{\bm{\beta}}))^2} \le \phi_{k}'(\bm{\beta}).
\end{align}
    \item Lower bound $l_{2,k}(\bm{\beta}, \bar{\bm{\beta}})$ of $\tau_{k}'(\bm{\beta})$: With \eqref{y} in Section \ref{SecW}, we have $\bm{y}_k(\bm{\bar\beta})=\bm{B}_{t}^{-1}(\bm{\bar\beta})\bm{a}_k(\bm{\bar\beta})$ and $\tau_{k}'(\bm{\beta})=\frac{1}{J}\bm{a}_k^{\text{H}}(\bm{\beta})\bm{B}_{t}^{-1}(\bm{\beta})\bm{a}_k(\bm{\beta})$. From \cite[Theorem 2]{FP}, we obtain
\begin{subequations}\label{l2}
  \begin{align}
  l_{2,k}(\bm{\beta}, \bar{\bm{\beta}})=&\frac{2}{J}\text{Re}\{\bm{y}_{k}^{\text{H}}(\bar{\bm{\beta}})\bm{a}_k(\bm{\beta})\}-\frac{1}{J}\bm{y}_{k}^{\text{H}}(\bar{\bm{\beta}})\bm{B}_k(\bm{\beta})\bm{y}_{k}(\bar{\bm{\beta}}) \nonumber\\
  =&\frac{2}{J}\text{Re}\{\bm{u}^{\text{H}}_{k}(\bar{\bm{\beta}})e^{j\bm{\beta}}\}- \frac{1}{J}(e^{j\bm{\beta}})^{\text{H}}\bm{U}_{k}(\bar{\bm{\beta}})e^{j\bm{\beta}}+C_{1,k}(\bar{\bm{\beta}}) \nonumber\\
  \le & \tau_{k}'(\bm{\beta}),
\end{align}
where
\begin{align}
    \bm{Y}_{k}(\bar{\bm{\beta}})&=\text{diag}\{\bm{y}_k(\bm{\bar{\beta}})\},\\
 \bm{u}_{k}(\bm{\bar\beta}) & = \bm{G}_{\text{r},k}^{\text{H}}\bm{W}_{k}\otimes \bm{I}_{M}\bm{y}_{k}(\bm{\bar\beta})-\sum_{k'\in \bigcup_{t'=t}^{T}\mathcal{G}_t}v_{k'}\bm{G}_{\text{r},k'}^{\text{H}}\bm{Y}_{k}(\bm{\bar\beta})\bm{W}_{k'}\bm{W}_{k'}^{\text{H}}\bm{G}_{\text{d},k'}^{\text{T}}(\bm{y}_{k}^{\text{T}}(\bm{\bar\beta}))^{\text{H}},\\
    \bm{U}_k(\bm{\bar\beta})&=\sum_{k'\in \bigcup_{t'=t}^{T}\mathcal{G}_t}v_{k'}\bm{G}_{\text{r},k'}^{\text{H}}\bm{Y}_{k}(\bm{\bar\beta})\bm{W}_{k'}\bm{W}_{k'}^{\text{H}}\bm{Y}_{k}^{\text{H}}(\bm{\bar\beta})\bm{G}_{\text{d},k'},\end{align}\begin{align}
    C_{1,k} (\bm{\bar\beta})&=2\text{Re}\{\bm{y}_{k}^{\text{H}}(\bar{\bm{\beta}})\bm{G}_{\text{d},k}\bm{w}_{k}\}-\sum_{k'\in \bigcup_{t'=t}^{T}\mathcal{G}_t}v_{k'}\bm{y}_{k}^{\text{H}}(\bm{\bar\beta}) \bm{G}_{\text{r},k'}\bm{W}_{k'}\bm{W}_{k'}^{\text{H}}\bm{G}_{\text{d},k'}^{\text{H}}\bm{Y}_{k}(\bm{\bar\beta})\nonumber \\   
    & \quad-\sigma^2\bm{y}_{k}^{\text{H}}(\bar{\bm{\beta}})\bm{y}_{k}(\bar{\bm{\beta}}).
\end{align}
\end{subequations}
    \item Lower bound $l_{3,k}(\bm{\beta}, \bar{\bm{\beta}})$ of $l_{2,k}(\bm{\beta}, \bar{\bm{\beta}})$: We  use the second order Taylor expansion as  the lower bound of $l_{2,k}(\bm{\beta}, \bar{\bm{\beta}})$:
\begin{align} \label{l3}
  l_{3,k}(\bm{\beta}, \bar{\bm{\beta}})=l_{2,k}(\bar{\bm{\beta}}, \bar{\bm{\beta}})+\nabla l_{2,k}(\bar{\bm{\beta}}, \bar{\bm{\beta}})^{\text{T}}(\bm{\beta}-\bar{\bm{\beta}}) - \frac{\kappa_k}{2}\|\bm{\beta}-\bar{\bm{\beta}}\|^2_2 \le l_{2,k}(\bm{\beta}, \bar{\bm{\beta}})
\end{align}
where $\nabla l_{2,k}(\bar{\bm{\beta}},\bar{\bm{\beta}})=2\text{Re}\{i\bm{\bar\theta}^{*}\odot (\bm{U}_k\bm{\bar{\theta}}-\bm{u}_k)\}$ is the gradient \cite{sca}, and $\kappa_k$ is a constant  on less than the Lipschitz constant of $\nabla l_{2,k}({\bm{\beta}},\bar{\bm{\beta}})$ \cite{mm}. Note that $\kappa_k$  can be chosen as follows.
The  gradient and the   Hessian matrix of $l_{2,k}({\bm{\beta}},\bar{\bm{\beta}})$ are respectively given by 
    \begin{subequations}
      \begin{align}
      \nabla l_{2,k}(\bm{\beta},\bar{\bm{\beta}}) & =2\text{Re}\left\{i (e^{i\bm{\beta}})^{*}\odot \left(\bm{U}_k(\bar{\bm{\beta}})e^{i\bm{\beta}}-\bm{u}_k(\bar{\bm{\beta}})\right)\right\},\\
          \nabla^2 l_{2,k}(\bm{\beta},\bar{\bm{\beta}}) &= 2\text{Re}\left\{
        \begin{bmatrix}
        (e^{i\beta_1})^*\bm{\alpha}_{k,1}\\
        \vdots\\
        (e^{i\beta_N})^*\bm{\alpha}_{k,N}
        \end{bmatrix}\right\},
      \end{align}
      where 
      \begin{align}
          \bm{\alpha}_{k,1} & =\left[ u_{k,1}-\sum\limits_{n\neq1} U_{k,1,n} e^{i\beta_n}  ,  U_{k,1,2}e^{i\beta_2} ,\cdots,U_{k,1,N}e^{i\beta_N} \right], \\
          \bm{\alpha}_{k,N} & =\left[U_{k, N,1}e^{i\beta_1} , U_{k,N,2}e^{i\beta_2},\cdots  , u_{k,N}-\sum\limits_{n\neq N} U_{k,N,n} e^{i\beta_n}  \right],
      \end{align}
      $u_{k,n}$ is the $n$th element of $\bm{u}_k(\bar{\bm{\beta}})$, and $U_{k,n, m}$ is the $(n, m)$th element of $\bm{U}_k(\bar{\bm{\beta}})$.
    \end{subequations}
    Note that $\|\nabla^2 l_{2,k}(\bm{\beta}, \bar{\bm{\beta}})\|_{\text{F}}$ is upper bounded by $\|\bm{\Gamma}_k\|_{\text{F}}$, where 
    \begin{align}
        \bm{\Gamma}_k=
        \begin{bmatrix}
        |u_{k,1}|+\sum\limits_{n\neq1} |U_{k,1,n}|  &  |U_{k,1,2}| &\cdots& |U_{k,1,N}|\\
        |U_{k,2,1}| &|u_{k,2}|+\sum\limits_{n\neq2} |U_{k,2,n}|  & \cdots& |U_{k,2,N}|\\
        \vdots & \vdots & \ddots & \vdots\\
        |U_{k,N,1}|   & |U_{k,N,2}| & \cdots & |u_{k,N}|+\sum\limits_{n\neq N} |U_{k,N,n}|
        \end{bmatrix}.
    \end{align}
 Then, we obtain
\begin{align}
    \| \nabla l_{2,k}(\bm{\beta}_1,\bar{\bm{\beta}}) -  \nabla l_{2,k}(\bm{\beta}_2,\bar{\bm{\beta}})\|_{2} \le \|\bm{\Gamma}_k\|_{\text{F}} \|\bm{\beta}_1-\bm{\beta}_2||_2,
\end{align}
which can be derived by using the mean value theorem for vector-valued functions. 
Thus,   $\nabla l_{2,k}(\bm{\beta},\bar{\bm{\beta}})$ is Lipschitz-continuous, where the Lipschitz constant $L$ satisfies $L\le\|\bm{\Gamma}_k\|_{\text{F}}$. By simply setting $\kappa_k= \|\bm{\Gamma}_k\|_{\text{F}}$, we obtain \eqref{l3} \cite{mm}.
\end{enumerate}

Combining \eqref{l1}, \eqref{l2} and \eqref{l3}, we obtain the lower bound of $\phi_{k}'(\bm{\beta}) - \rho_{\text{tar},k}$:
\begin{subequations}\label{lfunc2}
  \begin{align} 
  l_{k}(\bm{\beta}, \bar{\bm{\beta}})=-\frac{\alpha_{k}(\bar{\bm{\beta}})\kappa_k}{2}\|\bm{\beta}-\bar{\bm{\beta}}\|^2 +\alpha_{k}(\bar{\bm{\beta}}) \nabla l_{2,k}(\bar{\bm{\beta}},\bar{\bm{\beta}})^{\text{T}}(\bm{\beta}-\bar{\bm{\beta}}) +   C_{k}(\bar{\bm{\beta}}),
\end{align}
where
  \begin{align} 
  \alpha_{k}(\bar{\bm{\beta}})&=\frac{1}{J(1-\tau_{k}'(\bar{\bm{\beta}}))^2},\\
  C_{k}(\bar{\bm{\beta}})&=\phi_{k}'(\bar{\bm{\beta}})- J\alpha_{k}(\bar{\bm{\beta}})\tau_{k}'(\bar{\bm{\beta}}) + \alpha_{k}(\bar{\bm{\beta}})l_{2,k}(\bar{\bm{\beta}},\bar{\bm{\beta}})-\rho_{\text{tar},k}.
    \end{align}
\end{subequations}
Thus, Lemma 1 holds by letting $l_{2,k}(\bm{\beta}, \bar{\bm{\beta}}) = l_k'(\bm{\beta}, \bar{\bm{\beta}})$ in \eqref{lfunc2}.
\end{appendices}

\bibliographystyle{IEEEtran}
\bibliography{reference}

\end{document}